\documentclass[preprint]{aastex6}
\usepackage{graphicx}
\usepackage{rotating}
\usepackage{natbib}
\usepackage{aasref}
\usepackage{amsmath}
\usepackage{wrapfig}
\usepackage{placeins}
\usepackage{xcolor}

\begin{document}

\title{Spatial Variations in the Dust-to-Gas Ratio of Enceladus' Plume }
\author{M.M. Hedman$^a$, D. Dhingra$^{a,b}$, P.D. Nicholson$^c$, C.J. Hansen$^d$, G. Portyankina$^e$, S. Ye$^f$, Y. Dong$^e$}
\affil{$^a$Physics Department, University of Idaho, Moscow ID 83843 \\
$^b$ Department of Earth Sciences, Indian Institute of Technology Kanpur,
Kalyanpur, Kanpur 208016, Uttar Pradesh, India  \\
$^c$Department of Astronomy, Cornell University, Ithaca NY 14850 \\
$^d$Planetary Science Institute,  Tucson AZ 85719 \\
$^e$Laboratory for Astronomy and Space physics, University of Colorado, Boulder CO 80303 \\
$^f$ Department of Physics and Astronomy, University of Iowa, Iowa City IA 52242 
}
\begin{abstract}
On day 138 of 2010, the plume of dust and gas emerging from Enceladus' South Polar Terrain passed between the Sun and the Cassini spacecraft. This solar occultation enabled Cassini's Ultraviolet Imaging Spectrograph (UVIS) and the Visual and Infrared Mapping Spectrometer (VIMS) to obtain simultaneous measurements of the plume's gas and dust components along the same lines of sight. The UVIS measurements of the plume's gas content are described in Hansen {\it et al.} (2011, GRL 38:11202) , while this paper describes the VIMS data and the information they provide about the plume's particle content. Together, the VIMS and UVIS measurements reveal that the plume material above Baghdad and Damascus sulci has a dust-to-gas mass ratio that is roughly an order of magnitude higher than the material above Alexandria and Cairo sulci. Similar trends in the plume's dust-to-gas ratio are also found in data obtained when Cassini flew through the plume in 2009, during which time the Ion and Neutral Mass Spectrometer (INMS), Radio and Plasma Wave Science instrument (RPWS) and Cosmic Dust Analyzer (CDA) instruments made in-situ measurements of the plume's gas and dust densities (Dong {\it et al.} 2015 JGR 120:915-937). These and other previously-published systematic differences in the material erupting from different fissures  likely reflect variations in subsurface conditions across Encealdus' South Polar Terrain.
\end{abstract}

\maketitle

\section{Introduction}

The plume of vapor and small particles emerging from Enceladus'  South Polar Terrain is one of the Cassini Mission's most dramatic discoveries, and much has been learned about this moon's geological activity from the various instruments onboard the Cassini spacecraft. However, some basic aspects of this phenomenon remain obscure. In particular, the ratio of solid particles to molecular gases (primarily water vapor) in the plume is still not well constrained. This dust-to-gas ratio is an important parameter for understanding how particles and vapor are generated and accelerated beneath the moon's surface \citep{Porco06, Schmidt08, Kieffer09, IE11, Degruyter11, Yeoh15, Gao16}. However, it has been difficult to measure this quantity reliably because  the dust and vapor components of the plume are measured by different instruments under different conditions. Furthermore, the plume's properties vary with both space and time, complicating efforts to compare data obtained from multiple instruments.

 Estimates of the total vapor output of the plume have been derived from stellar occultations observed by the Ultraviolet Imaging Spectrograph (UVIS) and in-situ measurements obtained by the Ion and Neutral Mass Spectrometer (INMS) and Magnetometer during various close flybys. The occultation data yield estimates for total plume vapor output ranging between 180 kg/s and 250 kg/s \citep{Tian07,Hansen11, Hansen17}. By contrast, the in-situ measurements  have yielded more variable vapor outputs, ranging from 200 kg/s to around 1000 kg/s \citep{Saur08, Smith10, Dong11, Yeoh17}. While some of this range is due to different methods and assumptions used by different researchers, each individual analysis also finds that the total vapor output can vary by around a factor of two between different flybys. This apparent discrepancy may be because the in-situ measurements do not sample as large a region in the plume as the occultations do, but such explanations have not yet been evaluated quantitatively. The scale and magnitude of the temporal variations in the plume's gas output are therefore still rather uncertain.  
   
Meanwhile, the total particle mass output from Enceladus has been estimated to be around 50 kg/s based on an analysis of a few images obtained at extremely high phase angles \citep{IE11}, but recent work suggests that this mass could be a factor of several times lower if the particles are loose aggregates instead of solid grains \citep{Gao16}.  This mass estimate was also derived before systematic temporal variations in the plume's particle output  were documented. The brightness of the particle plumes varies by roughly a factor of four as Enceladus orbits Saturn, and its average brightness may even have decreased by 50-100\% over the course of the Cassini mission  \citep{Hedman13, Nimmo14, IE17, Porco17}. As with the potential variations in the plume's gas output, these changes in the particle output complicate any effort to quantify the dust-to-gas ratio in the plume.

Even more complexity is added to this problem because both the vapor and particle density are observed to possess noticeable spatial variations and structure \citep{Hansen11, Porco14, Spitale15, Dhingra17, Hansen17}. This not only complicates efforts to relate particle and vapor measurements made at different times and locations, but also hints that a single dust-to-gas ratio may not be the most useful parameter for characterizing Enceladus' geological activity. If nothing else, the observed variations in both time and space make simultaneous observations of both the particle and vapor components of the plume especially valuable.

The in-situ instruments onboard Cassini have been able to simultaneously measure the dust and gas densities in the plume during several close flybys of Enceladus \citep{Dong15}. These data confirm that both the dust and vapor components of the plumes can be highly structured, and also suggest that the mass density of solids in the plume (including sub-micron particles) is of order 20\% of the vapor density. These measurements probably provide the most reliable measurements of the dust-to-gas ratio to date. However, since Cassini can only sample a small fraction of the plume along a particular path during each flyby, extrapolating these findings to the entire average plume is challenging.
 
\begin{figure}
\resizebox{6in}{!}{\includegraphics{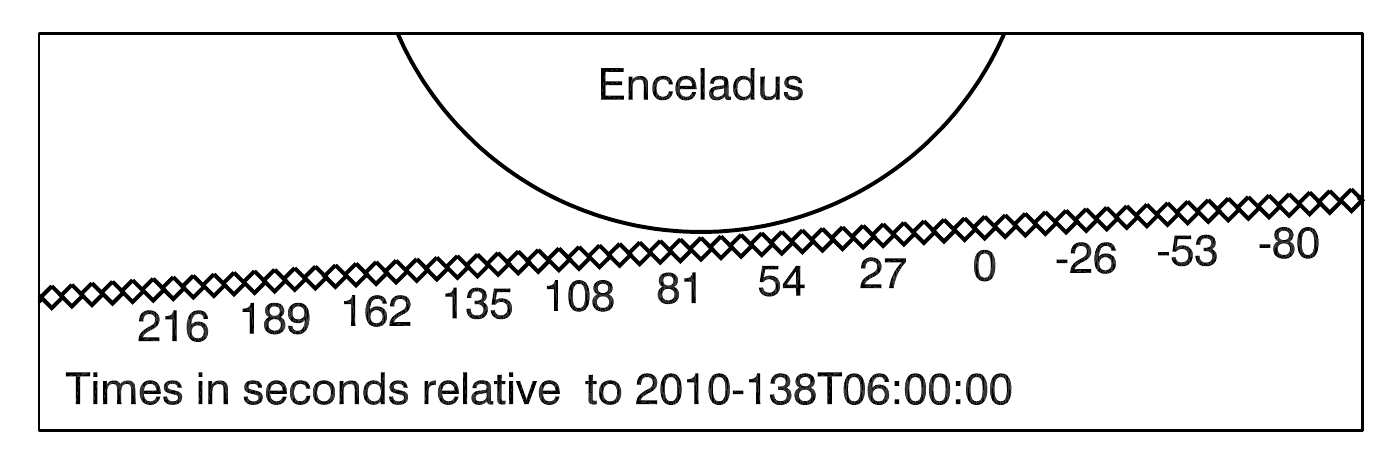}}
\caption{The path of the Sun relative to the disk of Enceladus as seen by the Cassini spacecraft on Day 138 of 2010. In this diagram, the projection of the moon's south pole points straight down. Note that the Sun never passes behind Enceladus itself, but it does pass behind the plume of material emerging from the moon's south pole (which extends below the moon in this view). Each diamond corresponds to the midtime of an individual VIMS measurement of the Sun's brightness.}
\label{geom}
\end{figure}

 A novel opportunity for Cassini to observe both components of the plume at the same time and place with remote sensing instrumentation occurred on day 138 of 2010, when the Sun passed just below Enceladus' south pole as seen by the spacecraft (see Figure~\ref{geom}). During this solar occultation, plume material should block some of the sunlight from reaching the spacecraft, and so both the Ultraviolet Imaging Spectrograph (UVIS) and the Visual and Infrared Mapping Spectrometer (VIMS) monitored the Sun's brightness as it passed by Enceladus. UVIS observed a clear decrease in the solar signal  at ultraviolet wavelengths that could be attributed to the plume's water vapor component \citep{Hansen11}. VIMS, by contrast, observed very little variation in the Sun's brightness  at near-infrared wavelengths (0.85 to 5.3 microns) during this same time. This was not surprising, since the water molecules that form the bulk of the plume's vapor component scatter infrared light much less efficiently than ultraviolet light. However, closer inspection of the VIMS data reveals a weak signal that appears to be due to the micron-sized particles in the plume. 
 
 If the signal observed by VIMS is really due to particles in the plume, then VIMS could constrain the total amount of solid material along the same lines of sight where UVIS measured the total amount of water vapor, and thus provide important information about the dust-to-gas ratio in the plume. Compared with the in-situ data considered by \citet{Dong15}, the VIMS and UVIS occultation data probe lower altitudes in the plumes because the line of sight to the Sun came within 20 km of the moon's surface. Furthermore, since material anywhere along the line of sight can scatter sunlight, the occultation data sample a broader region of the plume than the in-situ measurements. Intriguingly, the occultation observations reveal systematic spatial variations in the dust-to-gas ratio. Indeed, the dust-to-gas ratio of the material emerging from the different fissures appears to vary by roughly an order of magnitude.

Section~\ref{data} provides a brief overview of the solar occultation and the data obtained by VIMS, and how these data were processed to obtain estimates of the Sun's brightness as functions of time and wavelength. Section~\ref{signal} presents evidence that these data contain a feature that could be due to obscuration of the Sun by the plume, while Section~\ref{spec} shows that this signal has a spectrum consistent with scattering by plume particles. Section~\ref{comp} compares the VIMS and UVIS optical depth profiles to each other in order to constrain the magnitude and variations in the plume's dust-to-gas mass ratio. This section also compares these remote-sensing data with the in-situ measurements reported by \citet{Dong15}, demonstrating that the occultation data are broadly consistent with the in-situ data. Finally, Section~\ref{discuss} summarizes our findings and briefly discusses some of their implications for current efforts to understand what is going on beneath Enceladus' surface.

\begin{figure}
\centerline{\resizebox{4.in}{!}{\includegraphics{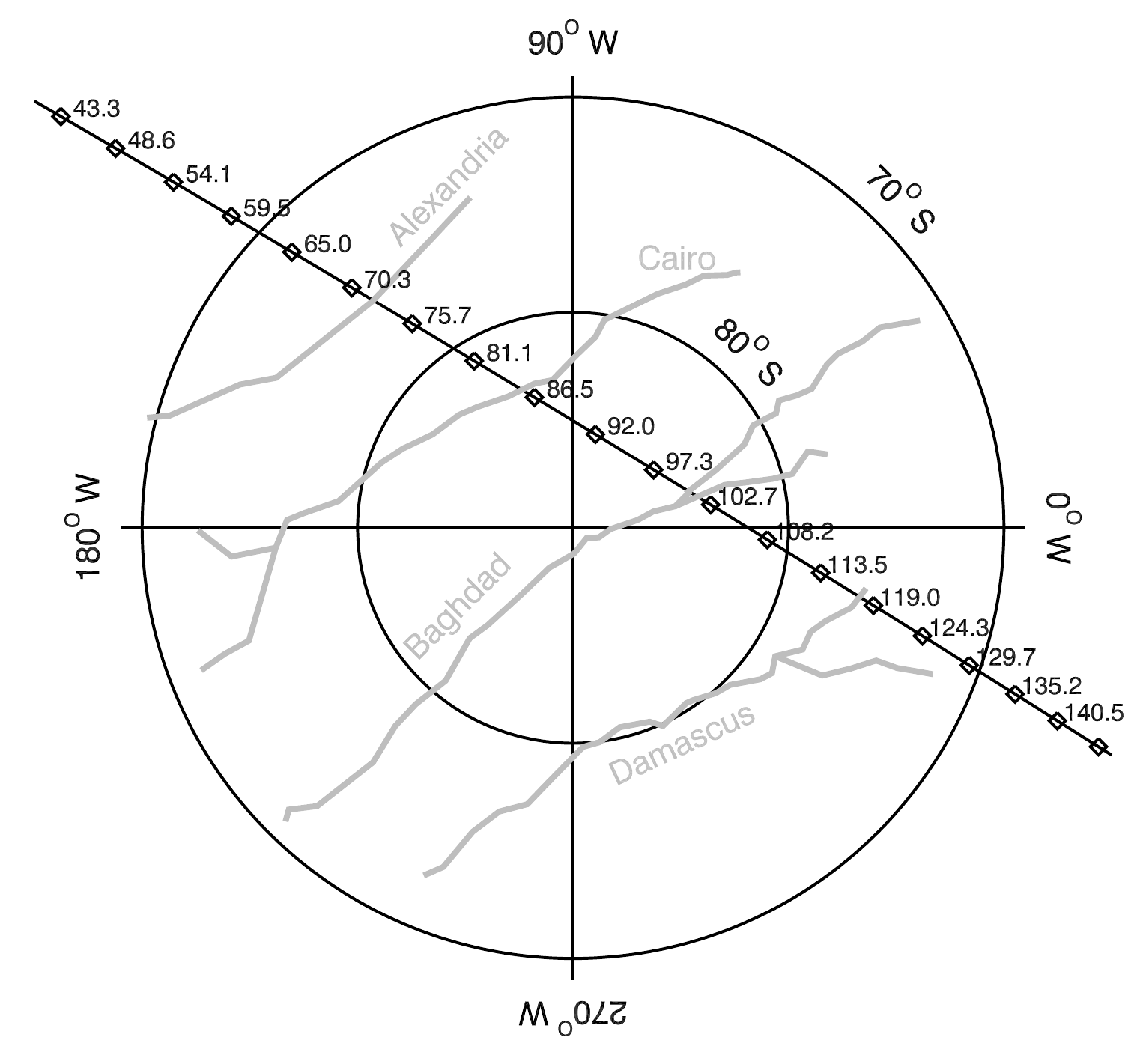}}}
\caption{The apparent trajectory of the solar occultation over Enceladus' south polar terrain. The line with diamonds shows the trajectory of the point along the line of sight towards the Sun that gets closest to Enceladus' center, projected vertically onto the moon's surface. The grey lines mark the locations of the large fissures where most of the plume material appears to be generated \citep{Porco14}. Each diamond marks the midtime of an individual VIMS measurement of the Sun's brightness.}
\label{geom2}
\end{figure}

\pagebreak 

\section{Observational Data}
\label{data}
\begin{figure}
\centerline{\resizebox{2in}{!}{\includegraphics{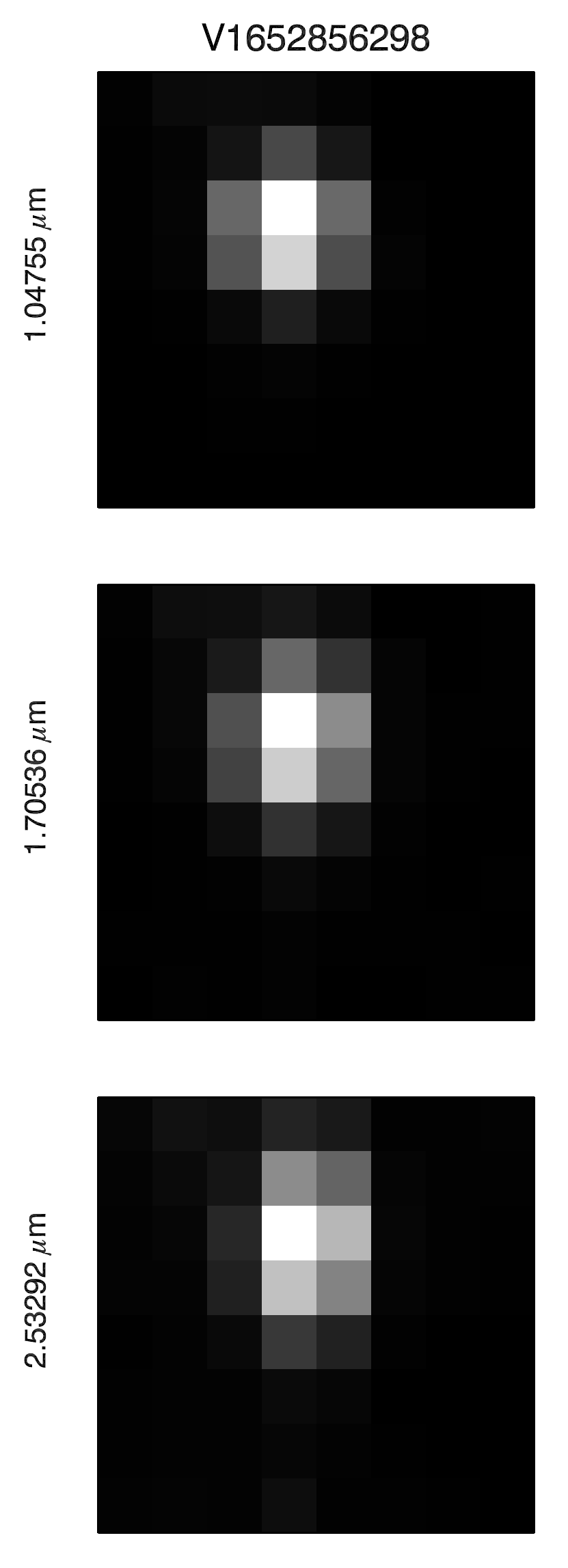}}}
\caption{The solar image recorded by VIMS. Each panel shows the brightness in a particular wavelength channel from the cube obtained closest to 2010-128T06:00:00. Each image has been separately stretched and shows a clear brightness peak due to the Sun. Note in these plots the VIMS $x$ and $z$ coordinates are horizontal and vertical, respectively.}
\label{ims}
\end{figure}

The UVIS data from this solar occultation have already been described and analyzed by \citet{Hansen11}, so this discussion will focus on the processing and analysis of the VIMS observations.  VIMS and UVIS each possesses a solar port that enables them to directly observe the Sun. For VIMS, this solar port is coupled via a separate optical path to the spectrometer, so that the instrument can obtain low-resolution images of the region around the Sun in 256 spectral channels between 0.85 and 5.2 microns \citep{Brown04}. These images are constructed by using scanning mirrors to observe a series of 0.5x0.5 mrad pixels that are packaged into three-dimensional ``cubes'' specifying the brightness of the scene as a function of two spatial dimensions and one spectral (wavelength) dimension.

During the solar occultation by Enceladus' plume, VIMS obtained a series of 308 cubes, each 8-by-8 spatial pixels across. The exposure duration for each pixel was 80 ms, which meant that each cube took about 5.4 seconds to complete. It took approximately 70 seconds for the Sun to pass over Enceladus' south polar terrain, so VIMS was able to obtain about 15 cubes during this critical time (see Figure~\ref{geom2}).  

This analysis uses uncalibrated VIMS data because the standard calibration pipelines do not account for the extra losses in the solar port optical path. This is not a major problem, however, because we are only interested in the fractional decrease in the Sun's brightness as it passed behind Enceladus' plume, and so we can use data obtained before and after the Sun passed behind the plume to normalize the data. Note that VIMS has a highly linear response function \citep{Brown04} and so the uncalibrated data numbers transmitted by the spacecraft are directly proportional to the apparent brightness of the scene.

Each cube has timestamps that enable the VIMS data to be correlated with the UVIS measurements and with the observation geometry. For each cube, we use the appropriate SPICE kernels \citep{Acton96} to determine the relative positions of the spacecraft, Enceladus and the Sun, and thus where the instrument's line of sight passed through the plume. For this study, useful geometrical parameters that can be derived from these data are the impact parameter of the line of sight (i.e. the minimum distance from the line of sight to Enceladus' center), as well as the latitude and longitude where the line of sight gets closest to Enceladus' surface (see Figure~\ref{geom2}).

Figure~\ref{ims} shows example images of the Sun obtained by VIMS during this occultation. Note that since the Sun is about 1 mrad across at Saturn, the solar image is only marginally resolved. Furthermore, the size and shape of the solar image varies slightly with wavelength due to the optics of the VIMS solar port. This peak is also superimposed on top of a finite background level that arises from scattered sunlight within the solar port. We therefore use multiple statistics to quantify the apparent brightness of the Sun at each wavelength in each cube.

One straightforward method of determining the total brightness of the Sun is to simply sum the data numbers from all 64 spatial pixels at each wavelength. However, as we will see below, this ``Total signal" is not ideal because it combines the direct signal from the Sun with the scattered light. While the scattered light also comes from the Sun, it follows a different optical path to the detector and so its level can be more dependent on the exact position of the Sun in the VIMS field of view. Indeed, as we will see below, the plume signal is not clearly detectable in this summed signal.

In order to isolate the direct signal from the Sun, we instead fit the solar images for every spectral channel in every cube to the two-dimensional gaussian peak function using the {\tt mpfit2dpeak} routine in IDL \citep{Markwardt09}. This program fits each image of the Sun to the following functional form:
\begin{equation}
{\rm Signal }= B+A e^{-(x-x_0)^2/2\sigma_x^2-(z-z_0)^2/2\sigma_z^2},
\end{equation}
where $x$ and $z$ are horizontal and vertical pixel coordinates, while $B$, $A$, $x_0$, $z_0$, $\sigma_x$ and $\sigma_z$ are fit parameters corresponding to the background level, peak amplitude, location and width, respectively.  These fits therefore yield estimates of all these parameters as functions of wavelength and time, which can then be used to estimate quantities like the Sun's apparent solid angle $\Omega_\Sun=2\pi \sigma_x\sigma_y$ and the total integrated signal under the peak $A\Omega_\Sun=2\pi\sigma_x\sigma_z A$. Note that this particular function assumes that the $x$ and $z$ axes are aligned with the major and minor axes of the gaussian. We verified that this assumption is valid for these particular  images by allowing the orientation of the gaussian axes to float (using the {\tt tilt} keyword in {\tt mpfit2dpeak}) and confirming that this did not alter the estimated amplitudes or solid angles of peak derived from the fit.

\begin{figure}[tb]
\resizebox{6.5in}{!}{\includegraphics{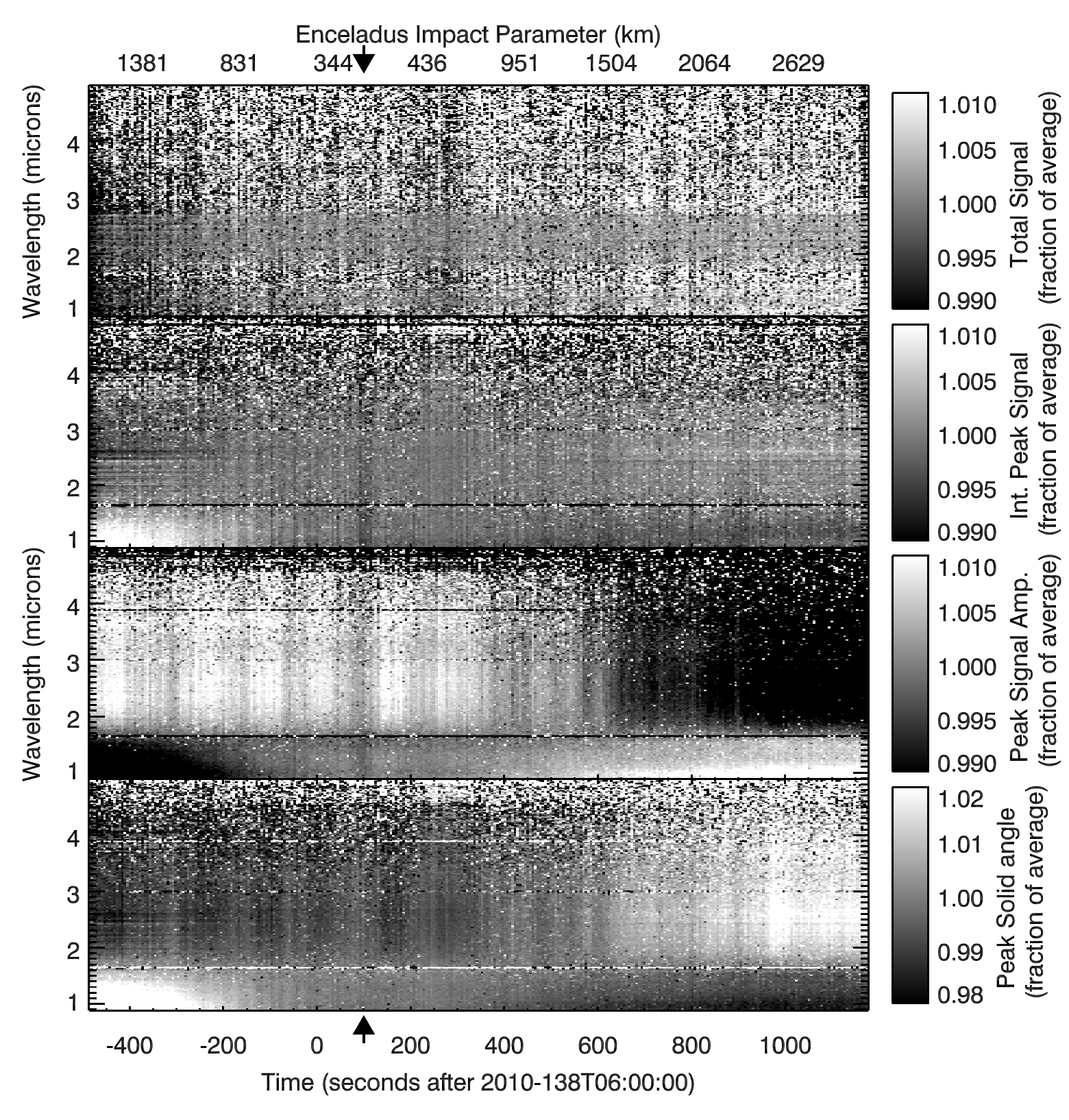}}
\caption{Overview of the VIMS Enceladus plume solar occultation data. Each panel shows the fractional variations in one observable parameter from the solar occultation as a function of wavelength and time. The top panel shows the total signal in the cube. The second panel shows the integrated brightness under the peak fit to the solar image. The third panel shows the amplitude of the peak derived from the fit to the solar image, and the bottom panel shows the solid angle of the solar image. The putative plume occultation signal corresponds to the vertical dark band in the second panel at around 100 seconds (marked with arrows on both axes), which occurred soon after the minimum impact parameter of the line-of-sight to the Sun.}
\label{overview}
\end{figure}

\begin{figure}[tbph]
\resizebox{6.5in}{!}{\includegraphics{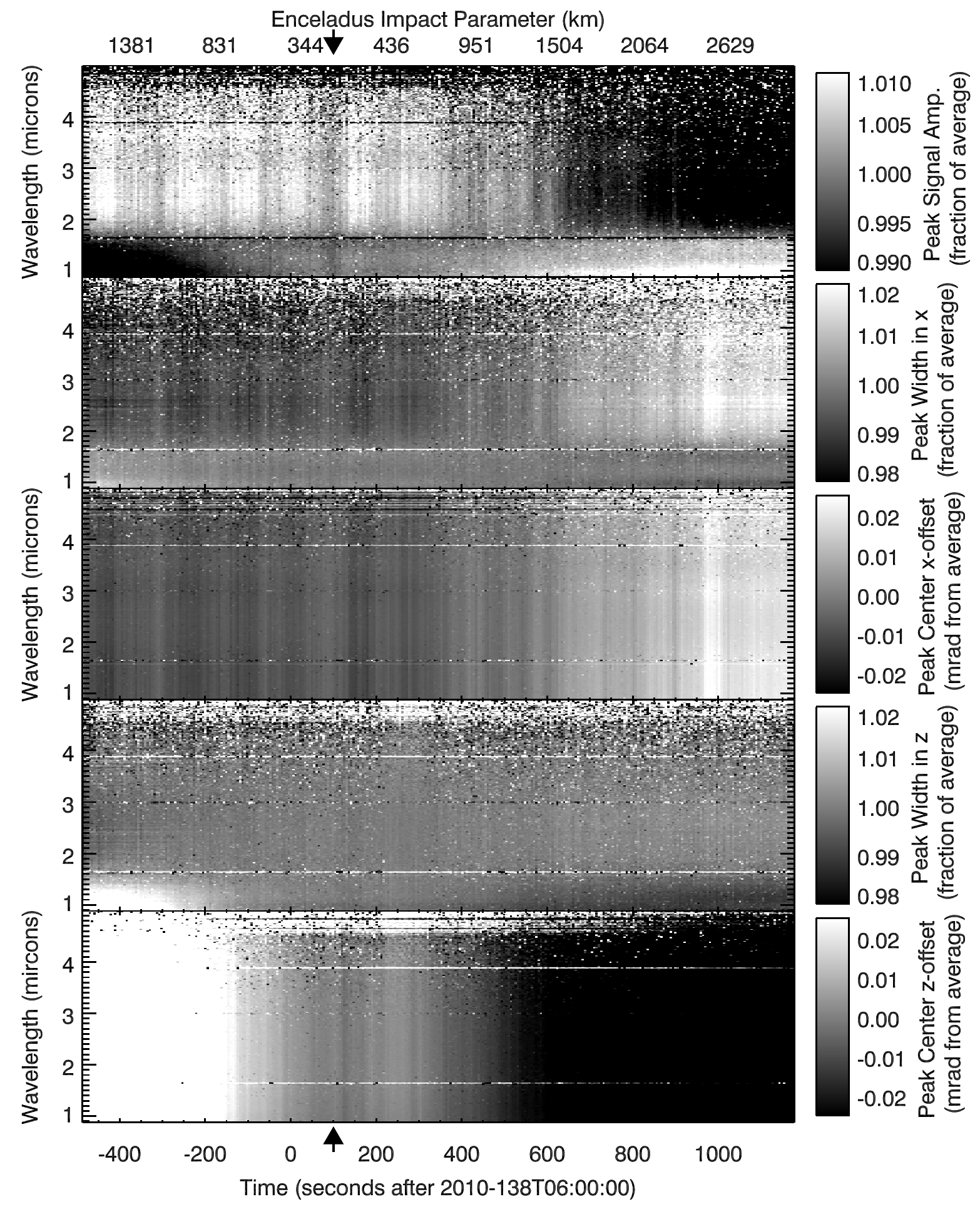}}
\caption{Individual parameters for the gaussian fits to the VIMS Enceladus plume solar occultation data. Like Figure~\ref{overview}, each panel show parameters as functions of wavelength and time. The panels show the amplitude, widths and locations of the peak. The amplitude and widths are normalized to their average values at each wavelength, and the positions are measured relative to the average values.}
\label{overview2}
\end{figure}

\begin{figure}[tbph]
\resizebox{6.5in}{!}{\includegraphics{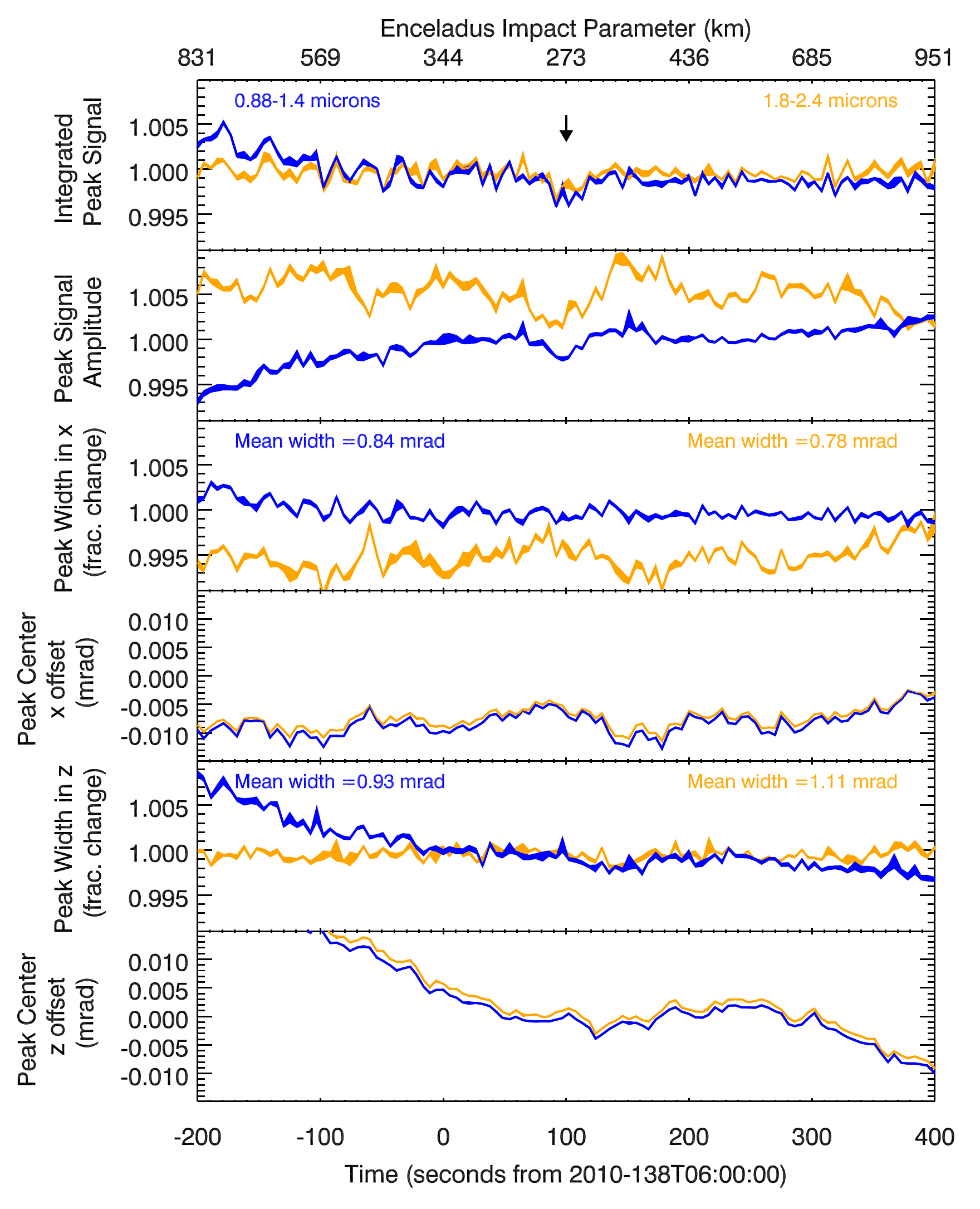}}
\caption{A closer look at the fit parameters in the VIMS Enceladus plume solar occultation data. Each plot shows the average values of selected parameters as functions of time. In each panel the two curves correspond to the average values of the parameter over the indicated range of wavelengths, and the widths of the line indicate the 1-$\sigma$ statistical uncertainty in these parameters based on the root-mean-squared variations among the relevant wavelength channels The top panel shows the signal integrated under the best-fit peak of the solar image, and the signal from the plume can be seen as the weak dip around 100 seconds (marked by the arrow). The other panels show the amplitude, widths and offsets in the peak. While the widths and positions do vary over this timeframe, none of these variations is obviously correlated with the dip in the integrated plume signal.}
\label{profiles}
\end{figure}

\section{Evidence for a plume signal in the VIMS data}
\label{signal}

Figure~\ref{overview} shows the fractional variations in four parameters related to the solar signal as functions of wavelength and time for the entire occultation observation. Specifically, it shows the variations in the total signal captured in the field of view, the total integrated signal under the solar image peak $A\Omega_\Sun$, the peak amplitude $A$ and the peak solid angle $\Omega_\Sun$.  The temporal variations in all of these parameters are of order 1\%,  which gives a sense of how stable these measurements are. However, we can also observe that at wavelengths less than 3 microns, the integrated signal under the peak $A\Omega_\Sun$ exhibits the smallest variations of all these parameters. The variations in this product are smaller than the variations in the two individual parameters because the variations in $A$ and $\Omega_\Sun$ are strongly anti-correlated with each other. However, the integrated peak signal is also more stable than the total signal in the image, probably because the  background level included in the latter involves non-standard optical paths that are more sensitive to the Sun's exact position in the solar port's field of view. The relative stability of the integrated peak signal suggests that it is probably the best way to quantify the solar flux.

Close inspection of the integrated peak signal in Figure~\ref{overview} reveals a dark band between 80 and 130 seconds after 06:00:00. This dip in signal occurs at the time when the spacecraft's line of sight passed closest to Enceladus over the south polar terrain (see top axis of Figure~\ref{overview}), and so it happened when we would be most likely to see obscuration of the Sun by the plume. Furthermore, this signal extends across a broad range of wavelengths, which is consistent with scattering by small dust grains (see below). Hence this dip is best candidate for actual obscuration of the solar signal by dust in Enceladus' plume. Even so,  this signal is quite small (the maximum obscuration being less than 1\%) and is not orders of magnitude larger than other variations in the signal level that may be due to instrumental phenomena (e.g. the slow decline in the  signal level over the course of the observation at wavelengths below 1.5 microns). Hence it is worth examining this dip in detail to verify that this is a real signal and not an instrumental artifact. 

The most straightforward way to assess whether this dip in the integrated peak signal could be due an instrumental artifact is to compare it with fluctuations seen in individual fit parameters like the peak amplitude $A$, peak widths $\sigma_x$ and $\sigma_z$ and peak offset positions $x_0$ and $z_0$. Figure~\ref{overview2} shows these parameters as functions of wavelength and time, while Figure~\ref{profiles} shows average profiles of these same parameters as functions of time for two different wavelength ranges in the vicinity of the putative plume signal. The width of the lines in the latter plot are equivalent to the statistical 1-$\sigma$ uncertainties in these parameters based on the root-mean-square variations in the parameters among the relevant wavelength channels. Note that these error estimates do not include systematic uncertainties associated with the instrument's response to time-variable phenomena, which are difficult to quantify {\it a priori}. Fortunately, the trends and correlations among the various parameters allow certain time-variations to be attributed to specific instrumental effects.

Both Figures~\ref{overview2} and~\ref{profiles} show that the apparent position of the Sun and the apparent width of the solar image do vary over the course of the observation. Close to the time of the putative feature, the $x$-position of the Sun oscillates  back and forth by a few microradians, while the $z$-position drifts more steadily towards more negative values at rates up to a few microradians per minute. The origins of these motions are still unclear, but they clearly affect the observed widths of the peak formed by the solar image. The $x$-axis position oscillations produce variations in the $x$-width of peak that are more obvious at longer wavelengths, and around 2 microns are about a few parts per thousand. By contrast, the drift in the $z$-axis centroid position seems to more strongly affect the $z$-width at shorter wavelengths, and around 1 microns we see trends of order a few parts in a thousand per minute. These correlations between peak positions and peak widths probably occur because VIMS just barely resolves the Sun and so slight changes in the Sun's apparent position impact how the signal is partitioned among different pixels. At the moment there is not enough information about the response function of the VIMS solar port to model these trends in detail, but the observed correlations are sufficiently clear to empirically document how small shifts in the Sun's apparent position impact the apparent size of the solar image.

The variations in the peak width are in turn anti-correlated with variations in the peak amplitude, which shows a combination of steady drifts and oscillations over the course of the observation. Again, these trends are sensible because as the Sun moves in the image, the solar image will be partitioned differently among the various pixels. Fortunately, these variations largely cancel out in the integrated signal, which is a measure of the total amount of light in the solar image (note that we obtained similar results by simply summing the signal above background in all the pixels containing the solar image). This cancelation is not perfect, however, and indeed in Figure~\ref{overview} we can see an overall trend at short wavelengths that tracks the variations in the $z$-offset and width, and some weak bands that could be associated with the stronger variations in the $x$-offset and width. 

Even though the integrated peak signal does show some sensitivity to these position variations, it does not appear that the dip around 100 seconds can be attributed to any of these features. As shown in Figure~\ref{profiles}, this dip does not have the same shape as the variations in either of the peak offsets. While the $x$-position does show a peak at 70 seconds and a dip at 150 seconds, which produced variations in both the peak amplitude and width at longer wavelengths, these variations are significantly broader than the dip in the integrated signal. Furthermore, around 1 microns the variations in the width are much smaller, and we can clearly see a dip in the amplitude that is comparable to the dip in the integrated signal at both wavelengths. The dip in the integrated signal therefore appears to be a distinct feature in these light curves that cannot simply be dismissed as an instrumental artifact, and so deserves serious consideration as a signal due to obscuration by the plume.

\section{Spectrum of the observed signal}
\label{spec}

To further explore whether this feature in the light curve is due to real obscuration of sunlight by the plume, we can examine how this signal varies with wavelength. Figures~\ref{overview}-~\ref{profiles} show that this signal extends over a broad range of wavelengths, which is consistent with light scattering by fine dust particles. However, Figure~\ref{profiles} also shows that  this signal is extremely weak, so we cannot expect to obtain reliable estimates of its spectral shape at different times. Instead, we characterize the signal's spectrum by computing its equivalent width as a function of wavelength. This quantity is proportional to the integrated area over the dip in the top panel of Figure~\ref{profiles}, and therefore provides an estimate of the total signal from the plume. In practice, we compute the equivalent width of the signal at each wavelength using the following procedure. First, we compute the integrated brightness under the peak for each cube at each wavelength to construct a light curve $A\Omega_\Sun$. These signals are then normalized to obtain estimates of the transmission $T(\lambda,t)$ by dividing $A\Omega_\Sun$ by an estimate of the unocculted solar signal. In practice, we determine this signal level by fitting $A\Omega_\Sun$ versus time on either side of the dip (-100 to +50 and +150 to +300 seconds relative to 2010-138T06:00:00) to a simple linear model. We then sum these transmission to get the equivalent width $EW$:
\begin{equation}
EW(\lambda)=\sum_{i=1}^N (1-T(\lambda, t_i)) v_i \delta t_i,
\end{equation}
where $\delta t_i$ is the time between adjacent cubes, and $v_i$ apparent speed of the Sun relative to Enceladus transverse to the line of sight.  In practice, we restrict the sum to cubes taken between 80 and 130 seconds after 06:00:00  because this is where the signal is clearly visible in the profiles and considering a larger timespan tended to reduce the signal-to-noise ratio. 
The uncertainty on this parameter is estimated using the following formula
\begin{equation}
\sigma_{EW}(\lambda) =\sigma_T(\lambda)\frac{1}{\sqrt{N}} \sum_{i=1}^N v_i \delta t_i,  
\end{equation}
where $\sigma_T(\lambda)$ is the standard deviation of the transmission estimates used to estimate the unocculted star signal.

\begin{figure}
\resizebox{6in}{!}{\includegraphics{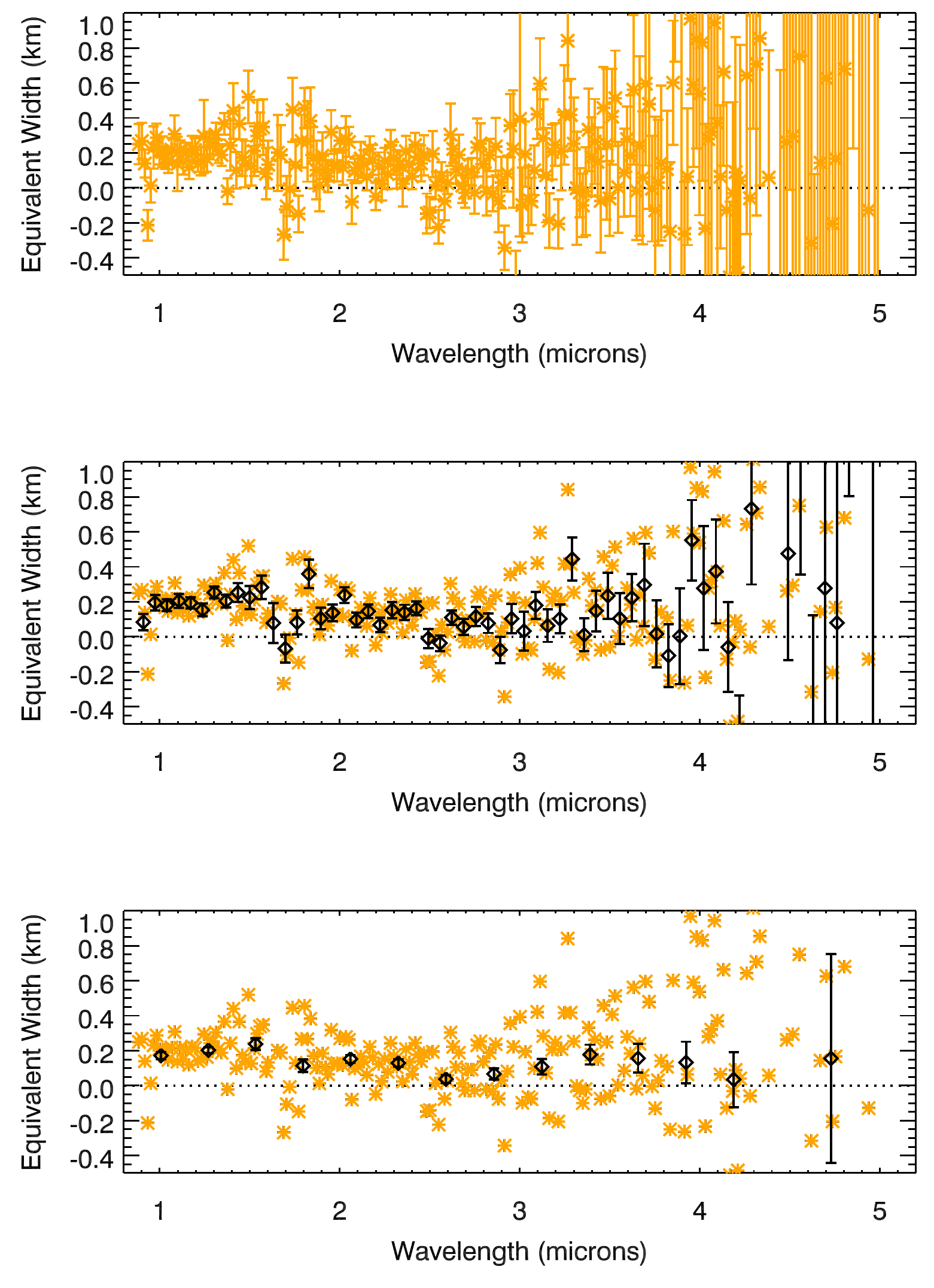}}
\caption{Spectrum of the candidate plume absorption signal. The top panel shows the equivalent width of the signal as a function of wavelength, with each data point corresponding to a single wavelength channel, and error bars corresponding to $1-\sigma$ uncertainties in the signal. In the lower two panels the stars show the same signal levels for individual channels as orange stars, while the black diamonds with error bars are averages over either four or sixteen spectral channels. For these averaged  spectra, there is a consistently positive signal between 1 and 3 microns. 
}
\label{specs}
\end{figure}

Figure~\ref{specs} shows the spectrum of these equivalent widths as a function of wavelength. The signal-to-noise clearly decreases with increasing wavelength, which is simply a result of the lower solar signal and higher instrumental thermal background. In addition, the signal level also appears to generally decline with increasing wavelength. The most obvious exception to this overall trend is a dip in the signal level around 1.7 microns. This feature falls close to a filter gap that is visible as a distinct horizontal band in Figure~\ref{overview},  and appears to correspond to a range of wavelengths where the amplitude is reduced over a broad timespan in Figure~\ref{overview2}, so we conclude that this feature is most likely an instrumental artifact. The somewhat weaker dip at 2.5 microns may have a similar origin.

Outside of these narrow features, the signal slowly decreases with increasing wavelength by roughly a factor of two between 1 and 3 microns (at longer wavelengths the uncertainties are  too large to discern any sensible trend). This trend is consistent with obscuration by small plume particles. Particles of size (radius) $s$ can only efficiently scatter radiation at wavelengths $\lambda$ shorter than $2\pi s$, and so outside of strong absorption bands, any population of fine particles will scatter and/or absorb short-wavelength radiation more efficiently than long-wavelength radiation \citep{vandeHulst}.

The shape of the observed  transmission spectrum is sensitive to the relative number of particles with different sizes, so in principle the observed trends could either be compared with expectations based on prior estimates of the plume's particle size distribution, or be used to provide an independent constraint on the plume's particle size distribution. In practice, like its overall brightness, the plume's particle size distribution varies with location and possibly time in ways that are not yet fully characterized \citep{Hedman09, IE11, Dong15, IE17, Dhingra17, Porco17}, so trying to predict the particle size distribution for this particular observation is challenging. Furthermore, these data do not have sufficient signal-to-noise over a broad enough range of wavelengths to discern any features in the size distribution like those found in reflectance data \citep{Hedman09, IE11}. 

Given these limitations, we will compare the data with the expected behavior of particle populations with size distributions that are pure power laws with power-law index $q$. That is, where the number of particles between size $s$ and $s+ds$ at a given time $t_i$ is:
\begin{equation}
N(s, t_i) ds =N_0(t_i) s^{-q} ds
\end{equation}
between particle sizes $s_{min}$ and $s_{max}$, and zero otherwise. While it is already clear that the plume particle size distribution is not a pure power law \citep{Hedman09, IE11, Dong15}, it turns out that for many plausible size distributions, the spectral trends shown in Figure~\ref{specs} are only sensitive to a limited range of particle sizes (see below). In these cases, the observables are insensitive to the endpoints $s_{min}$ and $s_{max}$, and we can regard $q$ as the effective power-law index of the size distribution in the vicinity of one specific particle size.


For a power-law size distribution, the optical depth $\tau$ of such a particle population is given by the integral:
\begin{equation}
\tau (\lambda,t_i) =\int_{s_{min}}^{s_{max}}  N_0(t_i) s^{-q} \pi s^2 Q_{ext}(\lambda,s) ds.
\end{equation}
where $Q_{ext}$ is a dimensionless extinction coefficient that depends on both the particle size and the optical constants of the material at the appropriate wavelength. So long as the optical depth $\tau << 1$ (which is the case throughout the plume), then $\tau$ approximately equals $1-T$, and so the equivalent width of the plume will be the appropriately weighted sum of optical depths. If $q, s_{max}$ and $s_{min}$ are all independent of time, then one can write the equivalent width as:
\begin{equation}
EW(\lambda)=\sum_{i=1}^N \tau(\lambda, t_i) v_i \delta t_i,
 =\int_{s_{min}}^{s_{max}}\mathcal{N}_0 s^{-q} \pi s^2 Q_{ext}(\lambda,s) ds.
 \label{EWpred}
\end{equation}
where $\mathcal{N}_0=\sum N_0(t_i) v_i\delta t_i$. Of course, if parameters like $q, s_{max}$ or $s_{min}$ vary over the course of the observation, the expression for the equivalent width becomes more complicated. However, even in such cases the above expression can provide a useful approximation of the plume's overall spectral behavior.

In general, $Q_{ext}$ is a function that goes to 0 when $s<<\lambda/2\pi$ and asymtotes to 2 when $s>>\lambda/2\pi$ \citep{vandeHulst}. While the exact shape of the transition between these two states depends on the particles' shapes and optical constants, to first order we can approximate $Q_{ext}$ as a step function, in which case the integral becomes (provided $q>3$, and assuming $s_{\min} << \lambda/2\pi <<s_{\max}$):
\begin{equation}
EW \simeq \int_{\lambda/2\pi}^\infty  \mathcal{N}_0 s^{-q} 2\pi s^2  ds =EW_0 \lambda^{3-q}.
\end{equation}
Thus the signal's spectrum should be close to a power law with an index $3-q$. Note that in the limit where $q>3$ and $s_{\min} << \lambda/2\pi <<s_{\max}$ this expression does not depend on either $s_{min}$ or $s_{max}$, which reflects the fact that observed spectrum at a given wavelength is most sensitive to particles with size $s \simeq \lambda/2\pi$. This also means that the spectral trends are relatively insensitive to any spatial variations in $s_{min}$ and $s_{max}$. Similarly, variations in $q$ with either particle size or spatial location will cause the equivalent width to deviate from the above form, but given the limited signal-to-noise and wavelength range available in these data, we can regard the $q$ of such a model as the average effective power-law index  for particles around 0.5 microns in radius. 

While the simple photometric model given above clarifies how the particle size distribution impacts the plume's transmission spectrum, it is also important to note that close to strong molecular absorption bands there can be substantial changes in the optical constants that can produce deviations from this overall trend. If we assume the particles are all spherical and composed of pure water ice, then $Q_{ext}$ can be computed analytically using Mie theory, and the integral evaluated for any chosen value of $q$. 

We fit the observed transmission spectrum to both a simple power-law model and to Mie-theory models using the IDL routine {\tt mie\_single.pro} (available from {\tt http://eodg.atm.ox.ac.uk}) and the optical constants for crystalline water ice reported by \citet{Mastrapa09}. For both model suites, we evaluate the $\chi^2$ of the misfit between the model and the data for a range of $q$ and identify the best-fit model as the one that yields the smallest value of $\chi^2$. In these calculations, we only consider spectral channels with equivalent width errors smaller than 0.5 km, since data with larger error bars did not significantly constrain $q$. We estimate the uncertainty on this value of $q$ by transforming the values of $\chi^2$ versus $q$ to probabilities to exceed and fitting the resulting peak to a Gaussian function. Note that for this calculation we scale the error bars such that the reduced $\chi^2$ of the best-fit model was unity. For the sake of comparison, we also fit these same data to a model where the equivalent width is independent of wavelength. 
 
\begin{figure}
\resizebox{6in}{!}{\includegraphics{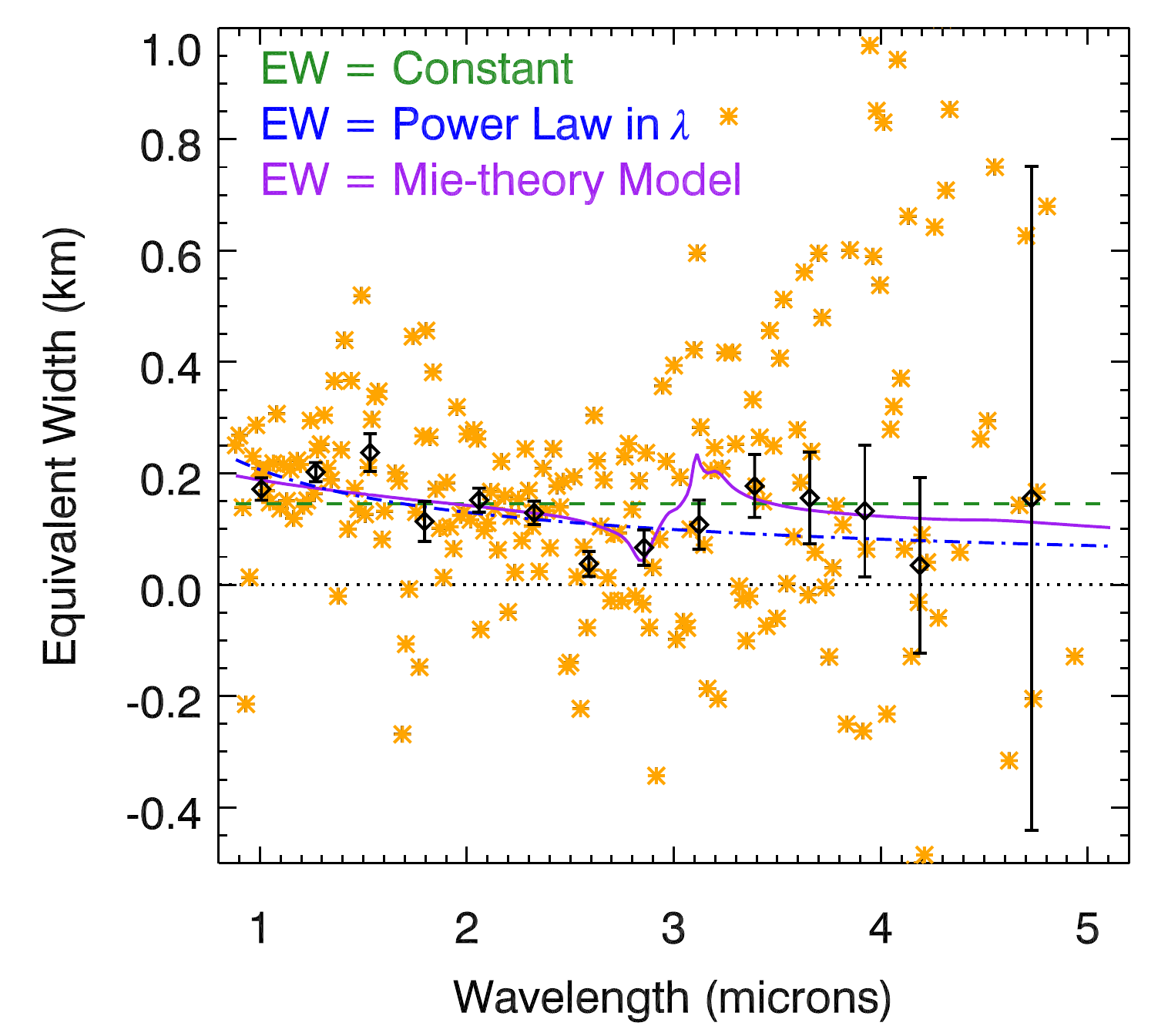}}
\caption{Comparing fits to the extinction spectrum. In both panels, the black and orange data points are the same as those shown in the bottom panel of Figure~\ref{specs}, while the three curves are simple model fits to the data (only to spectral channels with statistical uncertaintites less than 0.5 km). The green dashed curve is the best-fit uniform signal of $EW=0.14$ km. The blue dot-dashed curve show a simple power-law fit of the form $EW=$ (0.21  km)$\lambda^{-0.67}$. The purple curve is the best-fit Mie model (for a power-level size distribution with $q=3.1$). The difference in the quality of the latter two fits are not statistically significant.}
\label{spec2}
\end{figure}

 Figure~\ref{spec2} compares the observed spectral trends with these models. The wavelength-independent model is worse than the other two, (its $\chi^2$ being roughly 7\% larger), primarily because it systematically underpredicts the equivalent width at wavelengths shorter than 1.5 microns. For the simple power-law spectral models, the best-fit value for the particle-size power-law index $q=3.7\pm0.5$. Fitting the Mie-theory models yields a somewhat shallower size distribution with $q=3.1\pm0.6$. The difference in the index derived from these two models arises because the Mie-theory based models predict a fluctuation in the opacity around 3 microns due to the strong fundamental absorption band in water ice. These data have comparable $\chi^2$ values of 287 and 282 for 177 degrees of freedom, respectively. Note that these $\chi^2$ values are high, likely due to the narrow fluctuations around 1.7 and 2.5 microns that cannot be fit with these models. In spite of this limitation, it is clear that the data do not have sufficient signal-to-noise to discriminate between these two models and so we cannot claim a detection of any compositionally diagnostic spectral features.
 

These model fits indicate that the size distribution of the obscuring particles is slightly steeper than $s^{-3}$, which is consistent with previously published measurements of the plume's particle size distribution. \citet{Hedman09} fit high-phase reflectance spectra of the plume and found evidence that the size distribution of the plume became steeper than $s^{-3}$ for particles with radii greater than 1-2 microns. \citet{IE11} found models with a similar break could also fit brightness measurements obtained at very high phase angles. More recently, in-situ measurements by both the Cosmic Dust Analyzer (CDA) and the Radio and Plasma Wave Science instrument (RPWS) indicated that the plume's particle size distribution around 1 micron could be described by a power law with an index $q$  around $4$ \citep{Ye14}. Hence, even though the signal-to-noise is low, the signal in the solar occultation data does have a spectrum consistent with plume particles.

\section{Combining the VIMS and UVIS occultation profiles and comparing these data with in-situ measurements}
\label{comp}

Assuming that the signal seen by VIMS around 100 seconds after 2010-138T06:00:00 represents real obscuration by plume particles, we can compare this signal to that observed by the UVIS instrument \citep{Hansen11}. Since we are now primarily interested in the spatial variation in the plume signal, for this part of the analysis we computed the average transmission $T$ observed by VIMS at wavelengths between 0.85 and 1.8 microns,  and then converted this transmission to optical depth using the standard expression $\tau=-\ln(T)$. In practice, we estimated $T$ by first normalizing the solar signal at each wavelength to unity in two time periods that bracket the plume signal (one covers the time period between -100 and +50 seconds after 2010-138T06:00:00, while the other covers the time period between +150 and +300 seconds after 2010-138T06:00:00). The uncertainty for each of these transmission values was then estimated from the $rms$ fluctuations in the transmissions values in those same two time periods. Finally, we computed the weighted average of these individual transmission values to determine $T$ and used the corresponding error to determine the statistical uncertainty in this parameter, both of which could then be converted to optical depths. Note that since the signal-to-noise of these data are low, $\tau$ can be negative due to both noise fluctuations and instrumental artifacts that cause the signal to become more positive than the mean baseline level.

If the ratio of optical depths measured by UVIS and VIMS was a constant, then this number could be used to constrain the dust-to-gas mass ratio in the plume. However, it turns out that there are actually substantial variations in this optical depth ratio, which implies that the dust-to-gas ratio in the plume varies substantially across the plume. Hence, we first discuss these spatial variations in Section~\ref{spatial} and present evidence that they represent differences in the dust-to-gas ratio in the plume material above the different fissures. Section~\ref{ratio} then shows how we can use these data to estimate the absolute dust-to-gas mass ratios of the various plume regions. 

While this occultation provides a uniquely synchronous observation of the plume's dust and gas components for remote-sensing data, the various times the  Cassini spacecraft flew close to Enceladus' south polar terrain have enabled simultaneous measurements of dust and gas densities with various in-situ instruments \citep{Dong15}. More specifically, the data from the  ``E7" flyby on day 306 of 2009 is particularly useful because the spacecraft's ground track during this flyby is similar to the apparent motion of the Sun during the occultation. These in-situ data can therefore act as an independent check on both the spatial trends and the gas-to-dust mass ratios derived from the occultations. 

\begin{figure*}
\centerline{\resizebox{6in}{!}{\includegraphics{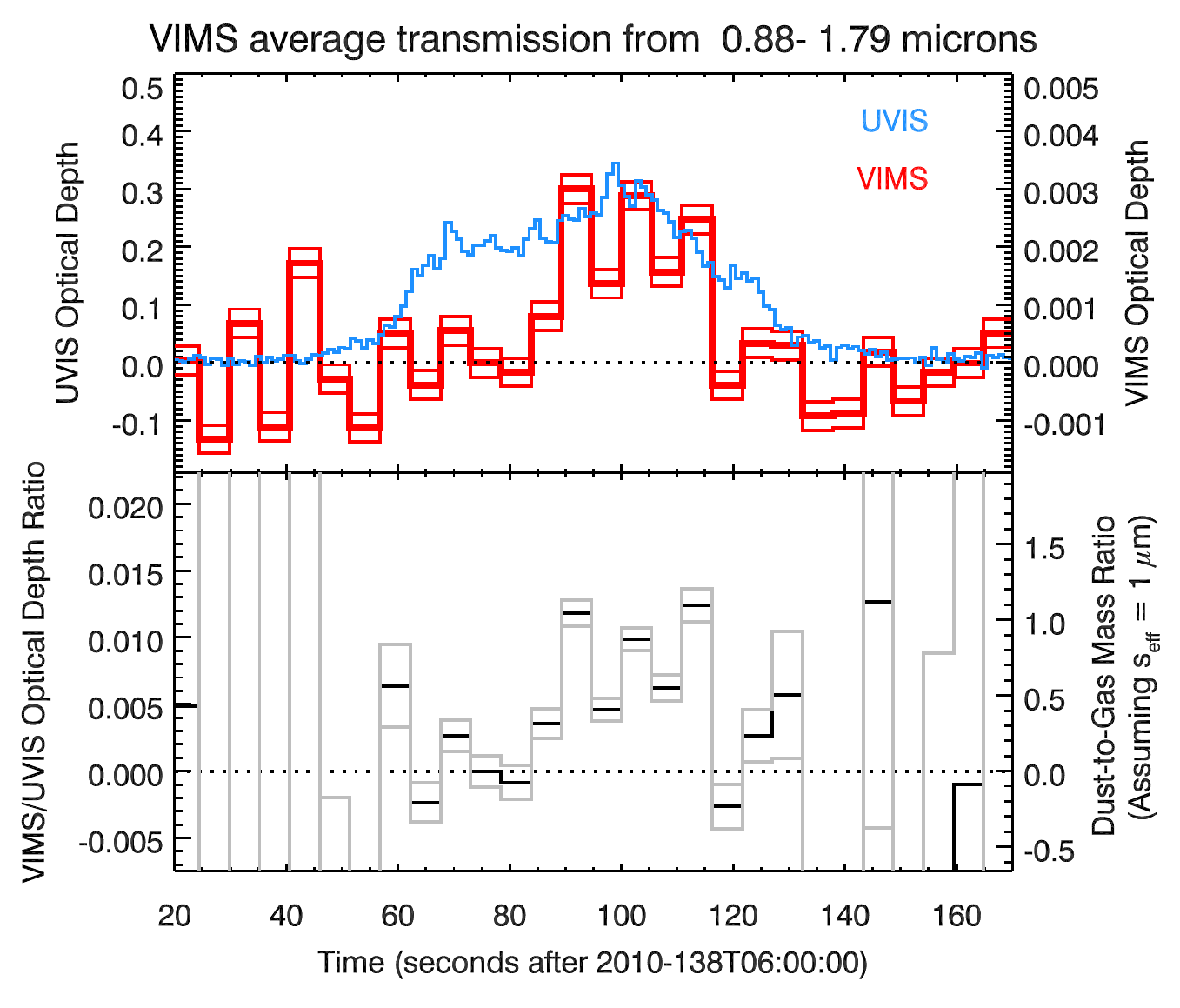}}}
\caption{Top: Optical depth profiles for UVIS and VIMS as functions of time. The VIMS profile is the average optical depth between 0.88 and 1.78 microns, and that the scales for the two profiles differ by a factor of 100, and the error bars on that profile are statistical uncertainties (see text).  Bottom: The ratio of optical depths measured by the two instruments, as well as the computed dust-to-gas column mass ratio assuming pure ice grains and an effective grain size of 1 micron (see Section~\ref{ratio}).}
\label{uvcomp}
\end{figure*}

\begin{figure*}
\resizebox{6in}{!}{\includegraphics{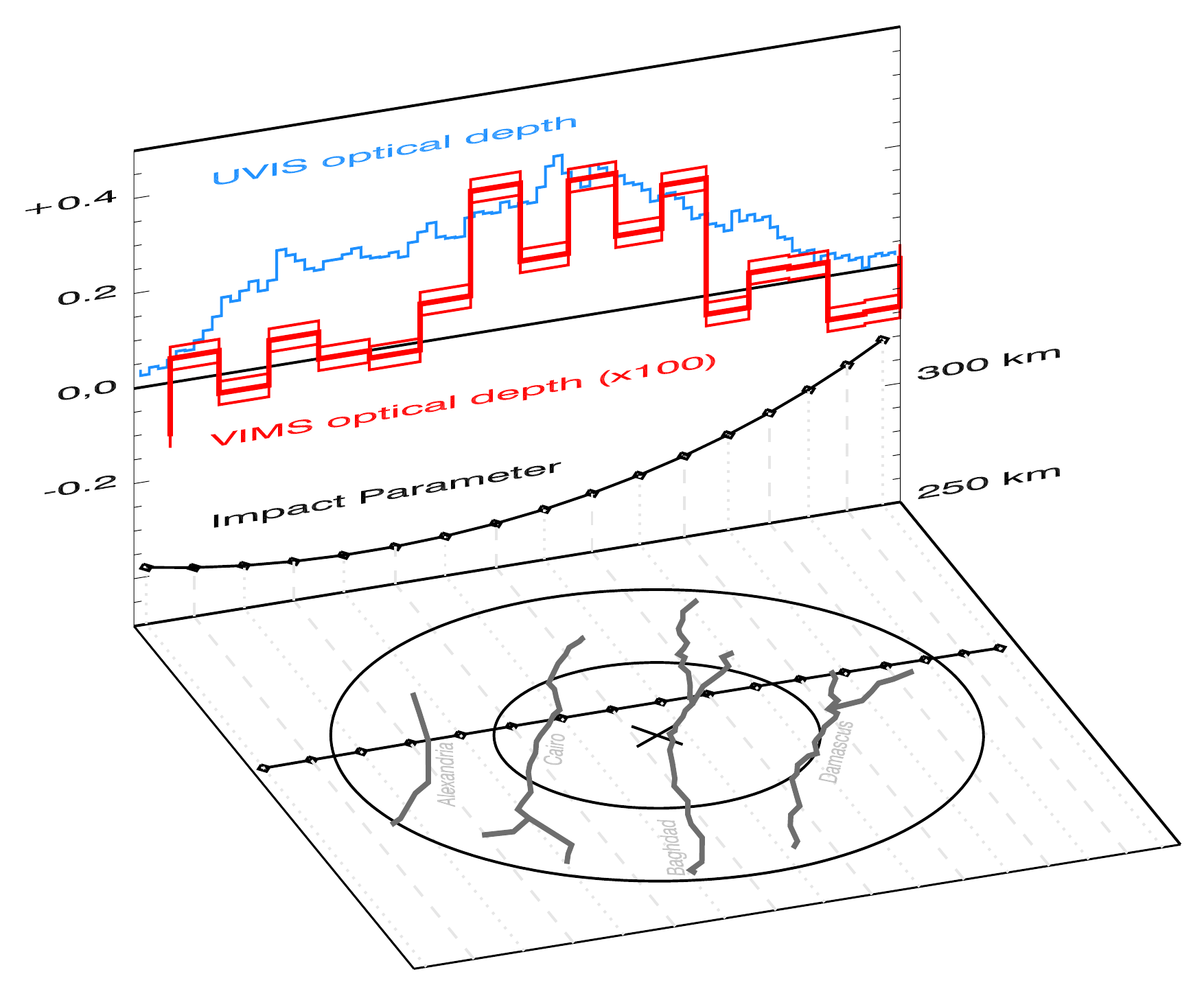}}
\caption{The context for the optical depth profiles from UVIS and VIMS. The bottom plane shows the south polar terrain, along with the ground track of the point where the line of sight towards the Sun was closest to Enceladus. The faint dotted and dashed lines correspond to projections of the approximate line of sight at the midtime for each VIMS cube. The back panel shows the two optical depth profiles, along with the impact parameter of the line of sight along this track (this is approximately 250 km more than the altitude above terrain). Note the VIMS signal is not strongest at lowest altitudes, but instead when the Sun was behind Baghdad sulcus.}
\label{uvcomp2}
\end{figure*}

\subsection{Spatial variations in the plume's dust-to-gas ratio}
\label{spatial}

The optical depth profiles recorded by the UVIS \citep{Hansen11} and VIMS (this work) instruments are shown on a common time axis in Figure~\ref{uvcomp}. Not surprisingly, the VIMS signal is much weaker than the UVIS signal. However, what is more interesting is that the two profiles have different shapes. UVIS observes a significant optical depth in the plume between 60 and 130 seconds after 2010-138T06:00:00, but VIMS only sees a clear signal between 85 and 115 seconds. Indeed, the ratio in the two optical depths appears to vary by almost an order of magnitude from less than 0.001 to around 0.01 across the plume.

There is no evidence that this difference between the UVIS and VIMS profiles is due to instrumental artifacts. As shown in Figure~\ref{profiles}, there is no fluctuation in any of the fit parameters that would explain a sudden change in the signal level around 85 seconds. Hence it appears that these differences reflect a real difference between the spatial distributions of particles and vapor within the plume. Indeed, since the optical depth ratio is directly proportional to the mean dust-to-gas mass ratio along the line of sight (see Section~\ref{ratio}), these data suggest that the dust-to-gas mass ratio could vary by roughly an order of magnitude.

The plume's spatial structure involves both vertical trends with altitude above Enceladus' surface and horizontal variations among the various fissures and sources within the South Polar Terrain. Strong altitudinal trends in the plume's dust-to-gas ratio are already reasonably well established. Observations with both ISS and VIMS reveal that a substantial fraction of the dust is launched at speeds below the moon's escape speed \citep{Porco06, Hedman09, Hedman13, Nimmo14, IE17, Porco17}. However, the vapor is emerging from the moon at substantially higher speeds \citep{Hansen06, Tian07}. The dust-to-gas ratio is therefore expected to steadily decline with increasing altitude. By contrast, horizontal trends in the plume's properties have been more difficult to quantify. Recent investigations have revealed systematic differences in the spectral properties of the material emerging from different fissures that likely represents variations in the particle size distributions launched from Cairo, Baghdad and Damascus sulci \citep{Dhingra17}, and hints of compositional variations in the plume particles emerging from different parts of the South Polar Terrain can be found in the published literature \citep{Brown06, Postberg11}. However, it is not clear how such trends in particle properties relate to the plume's gas content.

The relative importance of vertical and horizontal trends in the dust-to-gas ratio can be evaluated by placing the optical depth profiles in the context of the occultation geometry. Figure~\ref{uvcomp2} shows the Sun reached its minimum altitude {\em before} the VIMS signal transitioned from a low to a high level. Hence the low optical depth ratios occur at  lower altitudes above Enceladus' surface, which is inconsistent with the expected vertical trends for the plume's dust and gas. Instead, the observed variations appear to reflect variations in the dust-to-gas ratio among different plume sources across the fissures.

As shown in Figure~\ref{uvcomp2}, during this particular occultation the Sun passed behind material emerging from each of the tiger stripes in the order of Alexandria, Cairo, Baghdad and Damascus \citep{Hansen11}. The VIMS signal occurs when the Sun would be passing over Baghdad sulcus and part of Damascus sulcus, and so it appears that the material above those fissures has a higher dust-to-gas ratio than the material above Alexandria and Cairo sulci. Since we are observing the plume at finite altitudes, this could either mean that the material launched from Baghdad and Damascus is more dust-rich than the material launched from Cairo and Alexandria, or that a smaller fraction of the grains emerging from Cairo and Alexandria are launched at sufficient speed to reach the observed altitudes. While  the occultation data alone cannot distinguish between these two possibilities, other data sets could potentially discriminate between the two options. For example, this occultation probed the material emerging from Cairo and Alexandria at lower altitudes than the material emerging from Baghdad and Damascus. Hence, if the dust is only detectable above Baghdad and Damascus sulci because the material from those fractures reaches higher altitudes, then the dust launched from Baghdad  and Damascus must have a substantially larger scale height than the material launched from Cairo and Alexandria. Thus far no one has reported such dramatic differences in the scale heights of material emerging from the different fissures \citep[cf.][]{Porco14, Spitale15, Dhingra17}, which would probably be more consistent with the observed variations being due to differences in the total dust flux. A comprehensive analysis of multiple data sets will probably be needed to fully decide this matter, but regardless of whether the observed variations reflect differences in the total quantity or the typical launch speeds of particles, there are certainly significant variations in the properties of the material being launched from the different fissures.


\begin{figure*}
\centerline{\resizebox{6in}{!}{\includegraphics{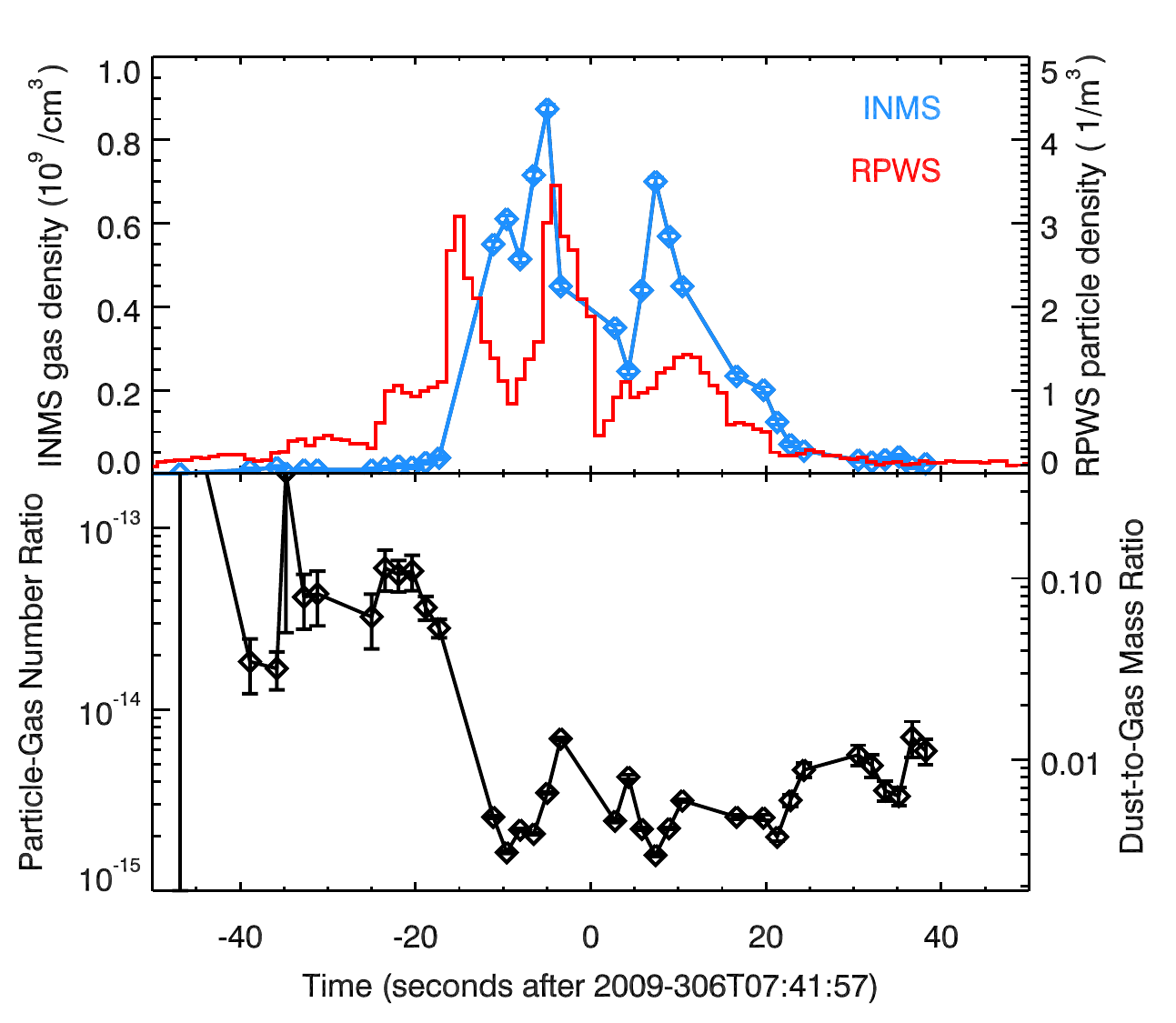}}}
\caption{Top: profiles of the gas and particle densities recorded by the INMS and RPWS instruments during the E7 flyby in 2009, adapted from \citep{Dong15}. The diamonds mark the times when the gas density was actually being measured by that instrument. Bottom: The ratio of particle to gas densities  measured by the two instruments, with the mass ratio computed assuming that the size threshold for RPWS is 1 microns, and that the size distribution spans at least two orders of magnitude (see Section~\ref{ratio}). Note the order-of magnitude variations in this ratio across the plume.}
\label{gdcomp}
\end{figure*}

\begin{figure*}
\resizebox{6in}{!}{\includegraphics{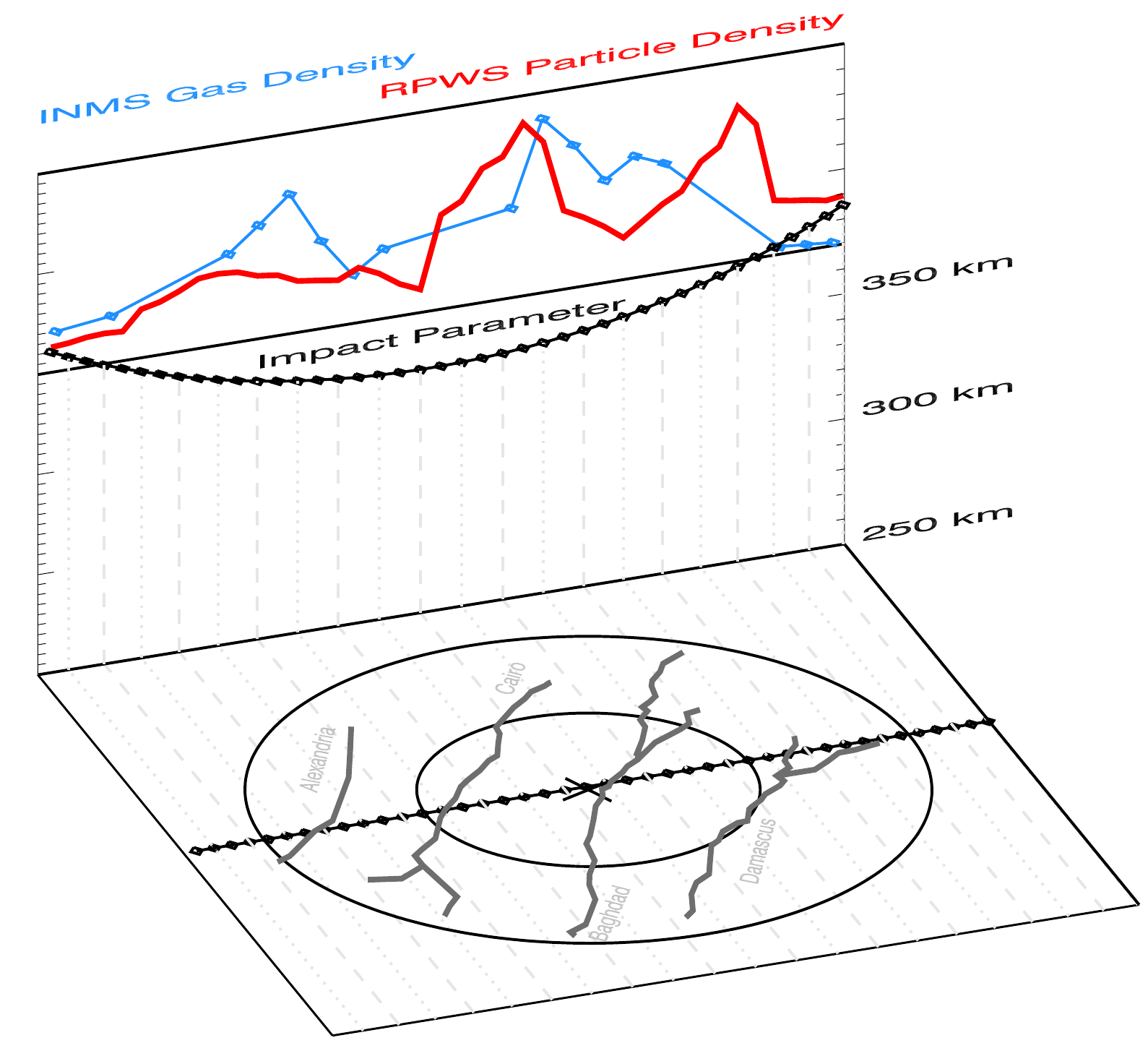}}
\caption{The context for the  INMS and RPWS data from the E7 flyby, shown in a similar format as Figure~\ref{uvcomp2}. Note the higher plume altitudes sampled during this observation, and recall that these two instruments sampled the plume along the trajectory shown in the two panels (the spacecraft moving from right to left over the course of the flyby). As with the UVIS/VIMS data, the higher dust-to-gas ratios appear to occur over Baghdad and Damascus, with the highest ratios occurring beyond Damascus.}
\label{gdcomp2}
\end{figure*}

Additional evidence for large  spatial variations in the plume's  dust-to-gas ratio can be found by examining trends in previously published in-situ measurements. Specifically, \citet{Dong15} compared estimates of the dust and gas density in the plume derived from the Ion and Neutral Mass Spectrometer (INMS) and Radio and Plasma Wave Science (RPWS) instruments from several close flybys through the plume. In particular, the E7 encounter in 2009 provides the most comparable dataset to the solar occultation because during this flyby the spacecraft flew horizontally across the various fissures in the South Polar Terrain, albeit in the opposite direction (i.e. Cassini flew over Damascus first, and Alexandria last during this flyby). Figure~\ref{gdcomp} shows estimates of the gas and particle densities as functions of time derived from those datasets. Both the INMS and RPWS data show a series of peaks that likely correspond to times when the spacecraft flew through different jets and sources. Unfortunately, INMS did not sample the first peak observed by RPWS, but one can note that the relative heights of the peaks at -5 seconds and +10 seconds differ by about a factor of two in the two datasets. Looking at the ratio of the two densities, we can also note that the dust-to-gas ratio is about an order of magnitude larger prior to the first peak than it is afterwards. The relatively high flux of dust at this time was observed by both the RPWS and CDA instruments \citep{Ye14}, so this cannot be attributed to a simple calibration error in RPWS. Hence these data appear to confirm that the dust-to-gas ratio can vary substantially across the plume.

Figure~\ref{gdcomp2} shows the E7 flyby data and geometry in a similar format as Figure~\ref{uvcomp2}. During the E7 flyby, the spacecraft crossed over the south polar terrain at a roughly constant altitude of around 100 km. As with the solar occultation data, the highest dust-to-gas ratios were not observed when the spacecraft was at its lowest altitude, and so vertical trends cannot explain the variations in the plume's dust-to-gas ratio. Instead, it again appears that material above Baghdad and Damascus sulci has higher dust-to-gas ratios than the material above Alexandria and Cairo. Indeed, the very highest dust-to-gas ratios were actually observed before the spacecraft flew over Damascus sulci. While the origin of this particle-rich plume material is still uncertain, the overall trends are reasonably consistent with those derived from the VIMS and UVIS occultation experiments.

\subsection{Estimating the dust-to-gas mass ratios in the plume}
\label{ratio}

Both the remote-sensing and the in-situ data demonstrate that the dust-to-gas ratio of the plume material varies dramatically across the plume, and these results are rather insensitive to issues associated with the calibration of the various instruments. These data can also be used to estimate the absolute values of the dust-to-gas mass ratios in different parts of the plumes. However,  
it is important to realize that such estimates are sensitive to assumptions about the particle size distribution, giving rise to substantial systematic uncertainties in the inferred dust-to-gas mass ratios. The following analysis provides explicit parameters in order to quantify these uncertainties that should help facilitate comparisons between these data and any theoretical predictions.

First, consider the VIMS and UVIS occultation data, where the observed optical depths can be converted into estimates of the total mass of particles and vapor along the observed lines of sight. For the following calculations, we will assume that the optical depth measured by UVIS ($\tau_{UVIS}$) is entirely due to water vapor molecules, while that measured by VIMS ($\tau_{VIMS}$) is entirely due to small ice particles. Also, we will assume that the optical depths are sufficiently low that they are proportional to the total cross-sectional area of all the molecules or particles in the plume. With these assumptions, we can convert the observed optical depths into estimates of the column mass densities in dust and gas using the following expressions:
\begin{equation}
M_{gas}=\left\langle\frac{m_{H_2O}}{\sigma_{H_2O}}\right\rangle \tau_{UVIS}
\end{equation}
\begin{equation}
M_{dust}=\left\langle\frac{m_{d}}{\sigma_{d}}\right\rangle \tau_{VIMS}
\end{equation}
where $m_{H_2O}$ and $\sigma_{H_2O}$ are the mass and cross-section of water molecules, while $m_d$ and $\sigma_{d}$ are the mass and cross sections of the ice particles in the plume. The angle brackets indicate either an average over wavelength (for UVIS) or an average over the full particle size distribution (for VIMS).

For the water vapor molecules, determining the relevant conversion factor is relatively straightforward. The optical depth profile plotted in Figure~\ref{uvcomp} is the average value between 850 and 1000 nm, and over this range of wavelengths the average cross section of a water molecule is $2\times10^{-17}$ cm$^2$ \citep{Chan93, Mota05}. Of course, the mass of an individual water molecule is 18 amu or  $3\times10^{-23}$ g. The relevant conversion equation therefore becomes:
\begin{equation}
M_{gas}=(1.5\times 10^{-6} g/cm^2) \tau_{UVIS}.
\label{gaseq}
\end{equation}

\begin{figure*}
\centerline{\resizebox{5.8in}{!}{\includegraphics{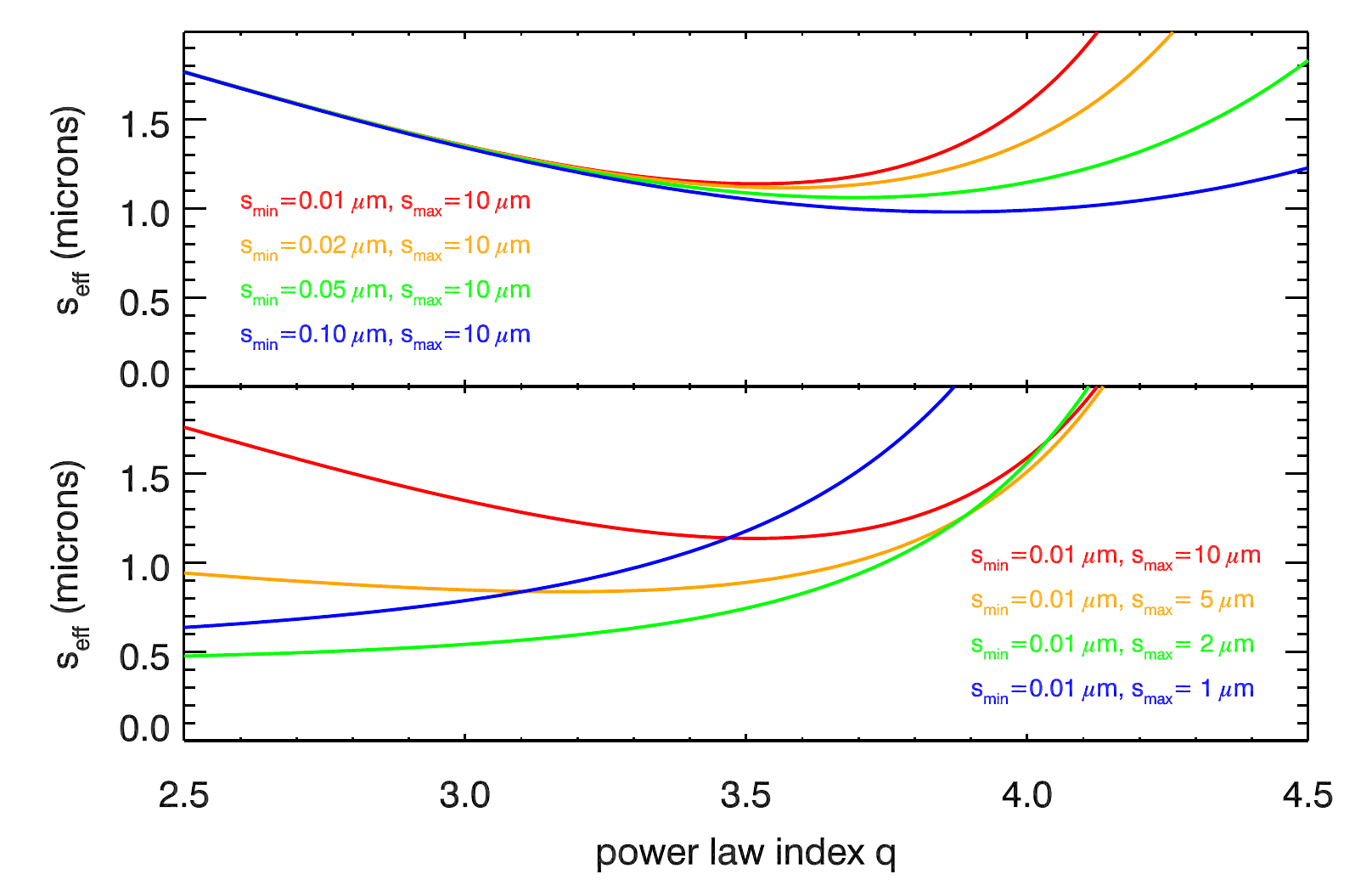}}}
\caption{Estimated values of the effective particle size for a range of different truncated power-law particle size distributions. In all cases the particle size distribution is assumed to be a truncated power-law, and $s_{\rm eff}$ is plotted as a function of the power law index $q$ assuming different values of the minimum (top panel) and maximum (bottom panel) particle sizes. In all cases the extinction coefficient is computed assuming a purely real index of refraction of 1.3, consistent with water ice.}
\label{seff}
\end{figure*}

\begin{figure*}
\centerline{\resizebox{5.8in}{!}{\includegraphics{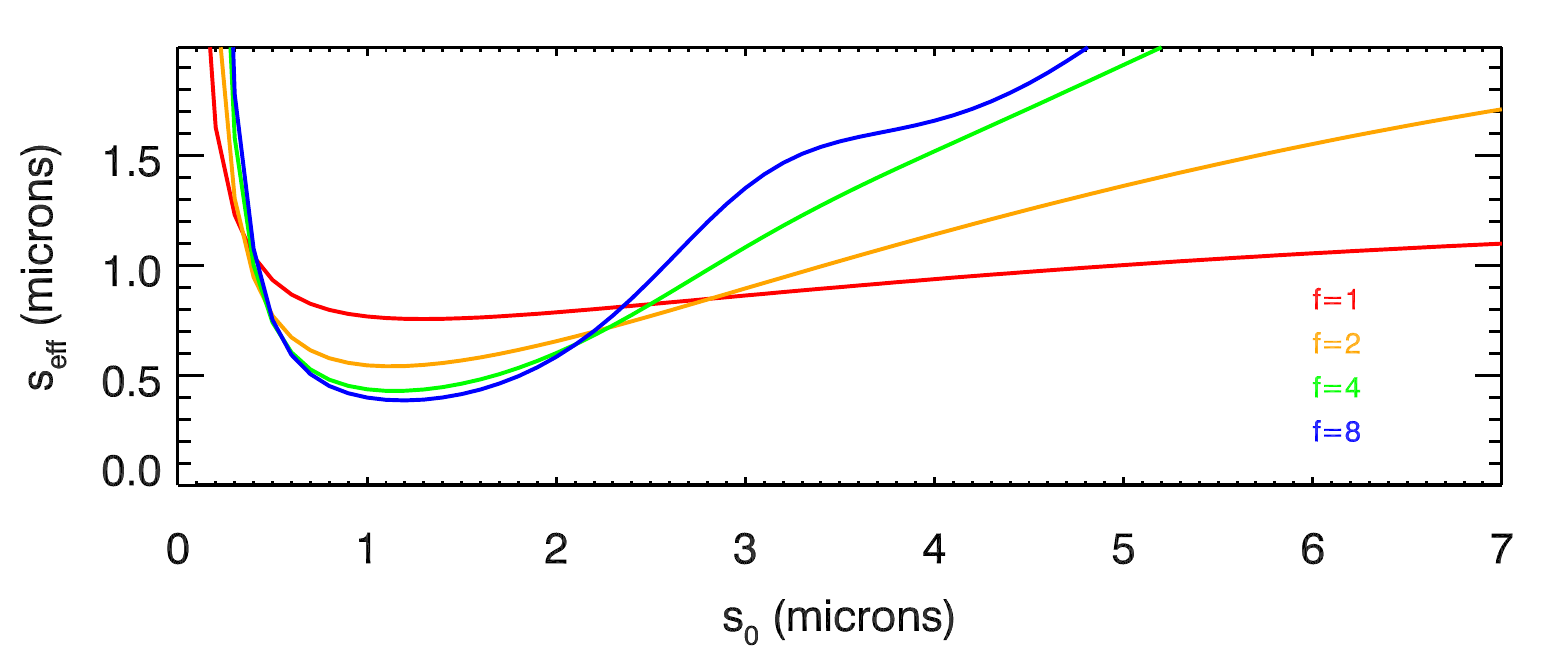}}}
\caption{Estimated values of the effective particle size for a range of particle size distributions of the form $N(s) \propto (s/s_0)^{f-4}/[1+(s/s_0)^{2f}]$ \citep{IE11}, with different values of the parameters $f$ and $s_0$. In all cases the extinction coefficient is computed assuming a purely real index of refraction of 1.3, consistent with water ice.}
\label{seff2}
\end{figure*}

For VIMS, the conversion factor is less straightforward because the plume particles have a range of sizes, and the conversion factor involves convolutions over the particle size distribution. For the sake of simplicity, we will here assume that all particles have the same average mass density $\rho$, but a range of sizes (radii) $s$. In this case, the conversion factor can be written in the following form:
\begin{equation}
\left\langle\frac{m_{d}}{\sigma_{d}}\right\rangle =\frac{\int \frac{4}{3}\pi\rho s^3 N(s)ds }{\int \pi s^2 Q_{ext} N(s) ds}=\frac{4}{3}\rho s_{\rm eff}
\end{equation}
where $s_{\rm eff}$ is an effective particle size given by the following expression:
\begin{equation}
s_{\rm eff}=\frac{\int s^3 N(s) ds}{\int Q_{ext} s^2 N(s) ds}
\end{equation}
These integrals can be evaluated for an arbitrary particle size distribution. Figure~\ref{seff} plots values of $s_{\rm eff}$ for a range of truncated power-law  size distributions with indices $q$ between 3 and 4 and a number of different maximum and minimum particle sizes $s_{\rm max}$ and $s_{\rm min}$. Menawhile, Figure~\ref{seff2} shows estimates of $s_{\rm eff}$ derived using more complex size distributions of the form $N(s) \propto (s/s_0)^{f-4}/[1+(s/s_0)^{2f}]$ used by \citet{IE11} for a range of different values of the parameters $f$ and $s_0$ (note that \citet{IE11} favored models with $s_0 \sim 3$ microns).  In all cases $Q_{ext}$ was computed using {\tt mie\_single} and assuming the particles' refractive index is 1.3 (appropriate for water-rich particles and wavelengths around 1.5 mictrons). For all of these situations, $s_{\rm eff}$ is of order 1 micron (i.e. comparable to the observed wavelength), so while this parameter does depend somewhat on the assumed particle size distribution, it is not exceptionally sensitive to uncertainties in the exact shape of the size distribution.

This means that we can write the relevant conversion factor in the following form:
\begin{equation}
M_{dust}=(1.3 \times 10^{-4} g/cm^2)\left(\frac{\rho}{1 g/cm^3}\right)\left(\frac{s_{\rm eff}}{1 \mu m}\right)\tau_{VIMS}.
\end{equation}
This can be combined with Equation~\ref{gaseq} to yield the following relationship between the optical depth ratio and the column mass ratio:
\begin{equation}
\frac{M_{dust}}{M_{gas}} = 120 \left(\frac{\rho}{1 g/cm^3}\right)\left(\frac{s_{\rm eff}}{1 \mu m}\right)\frac{\tau_{VIMS}}{\tau_{UVIS}}.
\end{equation}
Figure~\ref{uvcomp} shows the estimated dust-to-gas column mass ratio assuming a constant  effective particle size of 1 micron and an average particle density of 1 g/cm$^3$, which is reasonable for solid ice-rich grains. Of course, the variations in optical depth could potentially involve variations in the mass ratio, the effective particle size, or both. However, for these particular parameter values  the mass ratio is of order unity over Baghdad sulcus, indicating comparable masses of solids and vapor at these altitudes in the plume.

The reasonableness of these dust-to-gas ratios can be evaluated by comparing the occultation measurements with the in-situ data from the various plume flybys. \citet{Dong15} found solid-to-gas mass ratios of around 20\%, but comparing these numbers directly to ours is challenging because the \citet{Dong15} estimate of the plume's solid fraction included nanograins and the in-situ data were obtained at higher altitudes around the plume. A thorough comparison of the in-situ and occultation data would therefore require detailed examinations of the plume's vertical structure, as well as potential temporal variations in the moon's activity level. Such work is beyond the scope of this paper, and so we will instead simply compare the above estimate of the plume's dust-to-gas ratio with the most comparable in-situ data, which again comes from the E7 flyby.

During the E7 flyby INMS measured peak water vapor densities $\mathcal{N}_{INMS} = 10^{9}$ molecules/cm$^3$. At the same time, CDA and RPWS measured impact rates of particles larger than a threshold around 1 or 2 microns in radius depending on the ram speed and for RPWS, also the receiver gain. To eliminate the effects of receiver gain changes, the RPWS densities are scaled to a single threshold size of order 1 microns. The peak particle densities $\mathcal{N}_{RPWS}$ obtained in this way were a few particles per cubic meter \citep{Dong15, Ye14, Ye16}. As with the optical depths, these number densities can be converted to mass densities of gas and dust ($\mathcal{M}_{gas}$ and $\mathcal{M}_{dust}$, respectively) using the following formulae:
\begin{equation}
\mathcal{M}_{gas}=m_{\rm H_2 O} \mathcal{N}_{INMS}
\end{equation}
\begin{equation}
\mathcal{M}_{dust}=\langle m_p\rangle \mathcal{N}_{RPWS}
\end{equation}
where $m_{\rm H_2 O}=3\times10^{-23}$ g is again the mass of an individual water molecule, and $\langle m_p \rangle$ is the effective average mass of the particles detected by RPWS. The latter quantity can be written as the following ratio of integrals over the particle size distribution $N(s)$:
\begin{equation}
\langle m_p\rangle=\frac{\int_0^\infty \frac{4}{3}\pi \rho s^3 N(s)ds}{\int_{s_t}^\infty N(s) ds},
\end{equation}
where $\rho$ is the mass density of the grains and $s_t$ is the common threshold minimum particle size used to produce the density values shown in Figure~\ref{gdcomp}. Since we are averaging over $s^3$ instead of $s$, this expression is more sensitive to the assumed particle size distribution than the equivalent conversion factor for the occultations, and so for the sake of concreteness we will here assume that the relevant parts of the plume have a power-law particle size distribution with an index of 4 between the minimum and maximum particle sizes $s_{\rm min}$ and $s_{\rm max}$, and that $s_t$ is between these two size limits. These assumptions are consistent with the available in-situ observations of the micron-sized plume particles \citep{Ye14}, and allow the effective particle mass to be written as:
\begin{equation}
\langle m_p\rangle=4\pi\rho s_t^3 \ln(s_{\rm max}/s_{\rm min}),
\end{equation}
or  equivalently
\begin{equation}
\langle m_p\rangle=1.2\times10^{-11} g\left( \frac{\rho}{1 g/cm^3} \right)\left(\frac{s_t}{1 \mu m}\right)^3 \ln(s_{\rm max}/s_{\rm min}).
\end{equation}
Similarly, the dust-to-gas mass ratio can be written as:
\begin{equation}
 \frac{\mathcal{M}_{dust}}{\mathcal{M}_{gas}} =
4\times 10^{11}\frac{\mathcal{N}_{RPWS}}{\mathcal{N}_{INMS}} \\ \times \left(\frac{\rho}{1 g/cm^3} \right)\left(\frac{s_t}{1 \mu m}\right)^3 \ln(s_{\rm max}/s_{\rm min}).
\end{equation}
Note that in \citet{Ye14} and \citet{Dong15}, the threshold size $s_t$ was set to  2 microns, but more recent work by \citet{Ye16}  the data was recalibrated and $s_t$ reset to one micron.  Assuming $s_t=1$ micron, and that the particle sizes span at least two orders of magnitude (i.e. $s_{\rm max}/s_{\rm min} \simeq 100$), we obtain dust-to-gas mass ratios for the E7 flyby that range between about 1\% and 10\% (see Figure~\ref{gdcomp}). Note that residual uncertainties in the particle sizes recorded by RPWS could lead to order-of-magnitude systematic uncertainties in this mass ratio, but that  the strong variations in the dust-to-gas ratio across the plume are robust against such systematic uncertainties in the particle number density. 

Since the INMS/RPWS data were obtained at higher altitudes than the VIMS/UVIS data, it is not surprising that the dust-to-gas ratios would be lower for the in-situ measurements. Indeed, trends in the plume's brightness and optical depth near Enceladus' surface suggest that at low altitudes the particle and gas components have effective scale heights of around 30 km and 80 km, respectively \citep{Porco06, Hansen06}. One would therefore expect the dust-to-gas mass ratio to fall by roughly an order of magnitude between the 20-30 km altitudes probed by the occultations and the 100 km altitude probed by the E7 flyby. The dust-to-gas mass ratio derived from the occultation data therefore appears to be broadly consistent with the in-situ measurements. Of course, more detailed models that capture the full distribution of the particle and gas launch velocities will be needed to properly evaluate the consistency of these various measurements, but such analyses are well beyond the scope of this work.

We conclude this discussion of the plume's dust-to-gas mass ratio with two important cautionary points. First, all the above calculations assume that the particles are composed of solid ice grains. If the particles are instead loose aggregates, as suggested by \citet{Gao16}, the density of the particles could be well below 1 g/cm$^3$ and the dust-to-gas ratio will be correspondingly lower.  Second, even if the dust-to-gas mass ratio at 25 km altitude is of order unity, this does {\bf not} mean that the solid and gas {\em fluxes} are comparable to each other. The flux ratio also depends on the average velocities of the plume gases and particles, and the particles are known to be launched at substantially lower speeds than the water molecules \citep{Porco06,Hansen06, Tian07, Hedman09, IE17, Porco17}. Hence, even if the mass density of dust grains is as high as the mass density of vapor, the gas flux would likely be much larger than the particle flux from the surface.

\section{Discussion and Conclusions}
\label{discuss}

The VIMS and UVIS occultation data clearly demonstrate, and the independent RPWS/INMS data confirm, that the material emerging from different fissures have different characteristics, with the material emerging from Baghdad and Damascus sulci being more particle-rich than that emerging from Cairo and Alexandria sulci. This is yet another piece of evidence that there are systematic spatial variations in plume properties across Enceladus' South Polar Terrain. Recent examinations of the highest-resolution VIMS plume observations have documented spectral variations among the material emerging from different fissures that probably reflect variations in the corresponding particle size distributions, with Cairo material having distinctly different spectral properties from material erupting from Baghdad and Damascus  \citep{Dhingra17}. Furthermore, the published literature contains hints of systematic variations in the grain size of deposits around different fractures and in the composition of the grains lofted above the different fissures  \citep[Dhingra et al. {\it in prep}]{Brown06, Jaumann08, Postberg11}. These systematic spatial variations in source properties across Enceladus' South Polar Terrain likely reflect trends in the source material for the plumes and/or the plumbing connecting these reservoirs to the surface. Further explorations of what causes this spatial variability should therefore greatly enhance our understanding of Enceladus' geological activity and internal structure.

The occultation data also suggests that the dust-to-gas mass ratio of the material emerging from certain sources is close to unity at altitudes around 25 km. Since the particles are generally launched at lower velocities than the vapor, this suggests that even if the gas-flux from these sources is larger than the particle flux, the density of particles at the surface may be high enough to significantly affect the dynamics of the vent. These possibilities will need to be evaluated with more detailed simulations of both the particle and gas components of these systems.

\pagebreak

\section*{Acknowledgements}

The authors wish to thank the VIMS team, Cassini Project and NASA for the data used in this study. This work was supported in part by Cassini Data Analysis Program grant NNX15AQ67G. We also thank A. Ingersoll an anonymous reviewer for their comments on a previous draft of this manuscript.


\end{document}